\documentclass{aa}

\usepackage{graphicx}
\usepackage{wrapfig}
\usepackage{multirow}
\usepackage{txfonts}
\usepackage{lscape}
\usepackage{sidecap}   
\usepackage{adjustbox}
\usepackage{soul}
\usepackage{xcolor} 
\begin{document} 
   
   \title{Bar ages derived for the first time in nearby galaxies}

   \subtitle{Insights on secular evolution from the TIMER sample}
   
   \author{Camila de Sá-Freitas
          \inst{1}\thanks{\email{camila.desafreitas@eso.org}}
          \and Dimitri A. Gadotti\inst{2}
          \and Francesca Fragkoudi\inst{3}
          \and Paula Coelho\inst{4}
          \and Adriana de Lorenzo-Cáceres\inst{5,6}
          \and Jesús Falcón-Barroso\inst{5,6}
          \and Patricia Sánchez-Blázquez\inst{7}
          \and Taehyun Kim\inst{8}
          \and Jairo Mendez-Abreu\inst{5,6}
          \and Justus Neumann\inst{9}
          \and Miguel Querejeta\inst{10}
          \and Glenn van de Ven\inst{11}
         }

    \institute{European Southern Observatory, Alonso de Córdova 3107, Vitacura, Región Metropolitana, Chile
        \and Centre for Extragalactic Astronomy, Department of Physics, Durham University, South Road, Durham DH1 3LE, UK
        \and Institute for Computational Cosmology, Department of Physics, Durham University, South Road, Durham DH1 3LE, UK
        \and Universidade de São Paulo, Instituto de Astronomia, Geofísica e Ciências Atmosféricas, Rua do Matão 1226, 05508-090 São Paulo, SP, Brazil
        \and Universidad de La Laguna, Departamento de Astrofísica. Avda. Astrofísico Francisco Sánchez S/N, E 38206 La Laguna, Tenerife (Spain)
        \and Instituto de Astrofísica de Canarias. C/ Vía Láctea, S/N, E-38205 La Laguna, Tenerife (Spain)
        \and Instituto de Física de Partículas y del Cosmos (IPARCOS), Universidad Complutense de Madrid, 28040 Madrid, Spain
        \and Department of Astronomy and Atmospheric Sciences, Kyungpook National University, Daegu 41566, Republic of Korea
        \and Max Planck Institute for Astronomy, Königstuhl 17, 69117 Heidelberg, Germany
        \and Observatorio Astron{\'o}mico Nacional (IGN), C/ Alfonso XII 3, E-28014 Madrid, Spain
        \and Department of Astrophysics, University of Vienna, Türkenschanzstraße 17, 1180 Wien, Austria}

   \date{Received September 15, 1996; accepted March 16, 1997}

  \abstract{Once galaxies settle their discs and become self-gravitating, stellar bars can form, driving the subsequent evolution of their host galaxy. Determining the ages of bars can therefore shed light on the epoch of the onset of secular evolution. In this work, we apply the first broadly applicable methodology to derive bar ages to a sample of 20 nearby galaxies. The method is based on the co-eval build-up of nuclear structures and bars and involves using integral field spectroscopic (IFS) data from the Multi Unit Spectroscopic Explorer (MUSE) instrument on the Very Large Telescope to disentangle the star formation history of the nuclear disc from the background population. This allows us to derive the formation epoch of the nuclear disc and, thus, of the bar. We estimate the bar formation epoch of nearby galaxies -- mostly from the TIMER survey--, creating the largest sample of galaxies with known bar ages to date. We find bar formation epochs varying between $1$ and $13~\mathrm{Gyr}$ ago, illustrating how disc-settling and bar formation are processes that first took place in the early Universe and are still taking place in some galaxies. We infer the bar fraction over cosmological time with our sample, finding remarkable agreement with that obtained from direct studies of galaxies at high redshifts. Additionally, for the first time, we are able to investigate secular evolution processes taking into account the ages of bars. Our results agree with the scenario in which bars aid the quenching of the host galaxy, with galaxies hosting older bars tending to be more ``quenched''. We also find that older bars tend to be longer, stronger, and host larger nuclear discs. \textcolor{black}{Furthermore, we find evidence of the nuclear disc stellar mass build-up over time.} On the other hand, we find no evidence of downsizing playing a role in bar formation, since we find that bar age is independent of galaxy stellar mass. With the means to estimate bar ages, we can begin to understand better when and how bars shape the observed properties of disc galaxies.}
  

   \maketitle

%##################################################################%
%##################################################################%

\section{Introduction}

Studies suggest that present-day disc galaxies evolved in a two-phase scenario (e.g., \citealp{cook2010two}; \citealp{oser2010two}; \citealp{kraljic2012two}; \citealp{driver2013two}). At first, with a high incidence of mergers and interactions in the early Universe, the evolution of galaxies was led by external early events (e.g., \citealp{schreiber2006sinfoni}; \citealp{law2009kiloparsec}; \citealp{dekel2009formation}). This is reflected in characteristic turbulent morphologies and built dispersion-dominated structures. Later, with the expansion of the Universe, the rate of interactions and mergers decreased while, at the same time, galaxies assembled their masses. With that, they managed to dynamically settle into discs and internal processes took place, often referred to as secular evolution (e.g., \citealp{kormendy2004secular}). It is not clear, however, when discs settle in the Universe and when this transition occurred, a necessary benchmark for cosmological models.

To investigate when discs arise in the history of the Universe, we can analyse the morphology of galaxies at different redshifts. With the inauguration of the ALMA facilities, it became possible to probe the interstellar medium (ISM) at higher redshifts, revealing cold, rotationally supported discs up to $z\approx6$ (e.g., \citealp{smit2018rotation}; \citealp{neeleman2020cold}; \citealp{rizzo2020dynamically}; \citealp{lelli2021massive}; \citealp{posses2023structure}). More recently, JWST observations are finding that galaxies with exponential profiles, characteristic of disc galaxies,  exist at least since $z\approx8$, when the Universe was younger than $1~\mathrm{Gyr}$ old (e.g., \citealt{ferreira2022panic,ferreira2023jwst};  \citealp{nelson2022jwst}; \citealp{jacobs2023early}). Nevertheless, \cite{wang2024kinematic} demonstrated that galaxies can be photometrically flat with an exponential light profile and still be supported by random motion. Even though there is mounting evidence of the existence of discs at higher redshifts, the internal dynamics of said discs are still under debate. While some studies find that these discs are usually turbulent and thick (e.g., \citealp{elmegreen2006observations}; \citealp{cresci2009sins}; \citealp{newman2013sins}), other works find evidence for cold discs already at high redshifts (e.g. \citealt{rizzo2020dynamically}; \citealp{lelli2023cold}). Therefore, despite significant leaps forward over the last few years, the issue of when and how galactic discs settled in the Universe remains one of the most pressing questions in galaxy formation and evolution.

Theoretical works have proposed that bars form when the disc of its host galaxy is `dynamically mature' -- that is, self-gravitating with differential rotation, and rotationally supported with relatively low-velocity dispersion (e.g., \citealp{toomre1963distribution}; \citealp{hohl1971numerical}; \citealp{ostriker1973numerical}; \citealp{combes1981formation}; \citealp{gerin1990influence}; \citealp{combes1993bars}; \citealp{athanassoula2002bar}; \citealp{kraljic2012two}; \citealp{fragkoudi2024bar}). Therefore, the formation of a bar within its host galaxy will, in general, coincide or follow shortly after the formation of its dynamically cold stellar disc. Using cosmological zoom-in simulations, \cite{kraljic2012two} find that the bar formation takes place coincidentally with the disc maturing, specifically when the thin disc forms -- at least partially. Additionally, \cite{fragkoudi2024bar} find that an important factor for a galaxy to form a bar is when and if it becomes disc-dominated when compared to the dark matter halo. Recently, \cite{ghosh2023bars} demonstrated that bars can be formed in galaxies with both thin and thick discs. However, in the absence of a thin disc and with the thick disc presenting a large scale length and/or scale height, the bar formation is suppressed. This is in agreement with findings for the Milky Way as well, whose thin disc formed around $8~\mathrm{Gyr}$ ago (e.g., \citealp{haywood2013age}; \citealp{conroy2022birth}), coincident with the proposed time of bar formation of $7-8~\mathrm{Gyr}$ ago (e.g., \citealp{wylie2022milky}; \citealp{sanders2022mira}; \citealp{merrow2024did} -- however, see e.g., \citealp{nepal2024insights} and references within that propose young ages for the Milky Way bar).

Many studies have tried to understand when bars appeared first in the Universe. Looking back in time, different works measured the fraction of barred galaxies at different redshifts (e.g., \citealp{eskridge2000frequency}; \citealp{marinova2007characterizing}; \citealp{menendez2007near}; \citealp{sheth2008evolution}; \citealp{masters2011galaxy}; \citealp{simmons2014galaxy}; \citealp{melvin2014galaxy}; \citealp{le2024jwst}; \citealp{guo2024abundance}). These studies find, for the local Universe, fractions varying between $30\% - 70 \%$, depending on whether they include weak bars as well. These fractions decrease at higher redshifts, as shown by recent studies using JWST data -- \cite{le2024jwst} found bar fractions of $\sim14\%$ at $z\sim2-3$, and \cite{guo2024abundance} found $\sim 2.4-6.4\%$ at $z\sim3-4$. However, how much of this decrease is real is still under debate. That is because observational limitations such as spatial resolution, sensitivity and rest frame wavelength play an important role at higher redshifts. In this context, \cite{le2024jwst} compared bar fractions based on HST and JWST observations for the same sample, finding a difference factor of $\sim 2$. Additionally, cosmological simulations tend to give differing predictions on the fraction of bars across redshifts, with some simulations finding decreasing bar fractions (e.g., \citealp{kraljic2012two}; \citealp{fragkoudi2020chemodynamics}, \citeyear{fragkoudi2024bar}) while others find increasing bar fractions at higher redshifts (e.g. \citealp{rosas2020buildup}).

Another approach to understanding when bars appeared in the Universe is from galactic archaeology, i.e. to measure how old bars in nearby galaxies are. Until recently, however, we lacked a broadly applicable methodology to time bar formation in nearby galaxies and mostly had single-galaxy studies. Timing when bars are dynamically formed is not trivial and has been a challenge. This is because bars trap stars from the galactic disc, so the stars that compose the bar can be older than the bar itself. Instead, one can use bar-built nuclear structures, whose oldest stellar population will reflect the time of bar formation (e.g., \citealp{gadotti2015muse}). Theoretical works find that nuclear discs (aka pseudo-bulges or disc bulges) form shortly after bar formation -- $\sim10^8~\mathrm{yr}$ (e.g., \citealp{athanassoula1992existence},\citeyear{athanassoula1992morphology}; \citealp{lin2013hydrodynamical}; \citealp{emsellem2015interplay}; \citealp{fragkoudi2016close}; \citealp{seo2019effects}; \citealp{baba2020age}; \citealp{verwilghen2024simulating}). Once the bar forms, it induces shocks in the gas in the host galaxy disc, which thus loses angular momentum and funnels towards the centre of the galaxy, forming the nuclear disc. Therefore, one can use the moment of formation of the nuclear disc to indicate the formation of the bar. In \cite{de2023new}, we present the first broadly applicable methodology to measure bar ages in the nearby Universe, in a pilot study in which we measure the age of the bar in NGC\,1433 as $7.5^{+1.6}_{-1.1}~\mathrm{Gyr}$. In this work, we apply the same methodology to a larger sample of $18$ galaxies from the Time Inference with MUSE in Extragalactic Rings (TIMER -- \citealp{gadotti2019time}) project, deriving important insights on bar ageing and secular evolution for the first time from an observational perspective.

This paper is organized as follows: In Section \ref{sec_SampleDataDescr} we describe the properties of the TIMER sample and the observational data; in Section \ref{sec_Methodology}, we briefly describe the methodology presented in \cite{de2023new} and the modifications adopted in this work; in Section \ref{sec_results} we present the derived bar ages, and finally in Section \ref{sec_Discussion}, we discuss the implications of our results on disc settling, the bar fraction over time, bar formation in the downsizing context, multiple-barred systems, and bar-related secular evolution from an observation point of view for the first time. 
 
%##################################################################%
%##################################################################%

\section{Sample and data description }
\label{sec_SampleDataDescr}

Constraining bar ages for a large sample was not possible until recently with the methodology and pilot case presented in \cite{de2023new}. To apply the methodology presented in \cite{de2023new} to more galaxies, we need high spatial resolution integral-field spectroscopy from the central part of the galaxies, that is, from the region where the bar-built nuclear disc dominates. The TIMER survey (\citealp{gadotti2019time}) was specially designed with these characteristics and these goals. In this work, we apply our methodology to the TIMER sample to better understand when bars formed in nearby galaxies. 

TIMER is a high-quality data survey of the central region of $24$ nearby galaxies observed using the MUSE-VLT instrument -- from which $21$ were observed. The main goals of the survey include \textit{i}) estimating the cosmic epoch in which disc galaxies settled, hence forming the bar; and \textit{ii}) test if the downsizing scenario applies to the formation of bars \citep[see discussion in, e.g.,][]{sheth2012hot}. The sample was selected from the \textit{Spitzer} Survey of Stellar Structure in Galaxies (S$^4$G -- \citealp{sheth2010spitzer}) considering several morphological and observable characteristics. The TIMER survey focuses on massive galaxies ($\sim10^{10}-10^{11}~\mathrm{M}_\odot$), with visible bars and inner structures. The sample selection also imposes a limit on the distance (d$<40$ Mpc) and the inclination of the galaxies ($i<60^\circ$) to distinguish the different central structures, which also need to be observable from the Paranal Observatory, that is, with DEC~$< + 25^\circ$. Considering these constraints, the TIMER survey consists of $24$ galaxies, with distances within $40~\mathrm{Mpc}$.

\renewcommand{\arraystretch}{1.2}
\begin{table*}
\caption{Sample of galaxies used in this work}
\label{table_sample}
\centering
\begin{tabular}{lcccccccc}
 \textbf{Galaxy} & \textbf{Type} & \textbf{dist} & \textbf{inc} & \textbf{M$\star$} & \textbf{M$_{\mathrm{HI}}$} & \textbf{R$_{25.5}$} & \textbf{R$_\mathrm{kin}$} & \textbf{R$_\mathrm{bar}$} \\ 
  &  & [Mpc] & [deg] & [10$^{10}$ M$_\odot$] & [10$^{10}$ M$_\odot$] & [Kpc] & [Kpc] & [Kpc] \\
  & (1) & (2) & (3) & (4) & (5) & (6) & (7) & (8)  \\
 \hline \hline
 IC 1438  & R$_1$SAB$_\mathrm{a}$(r´\underline{l},nl)0/a & 33.8 & 24 & 3.1 & 0.12 & 12.7 & 0.60 & 4.15 \\
 NGC 613  & SB(\underline{r}s,bl,nr)b & 25.1 & 39 & 12.2 & 0.47 & 23.4 & 0.59 & 9.40 \\
 NGC 1097 & (R´)SB(rs,bl,nr)ab pec & 20.0 & 51 & 17.4 & 0.91 & 35.1 & 1.07 & 9.5 \\
 NGC 1300 & (R)SB(s,bl,nrl)b & 18.0 & 26 & 3.8 & 0.22 & 17.9 & 0.33 & 7.1 \\
 NGC 1433 & (R´$_1$)SB(r,p,nrl,nb)a & 10.0 & 34 & 2.0 & 0.07 & 12.7 & 0.38 & 3.66 \\
 NGC 3351 & (R´)SB(r,bl,nr)a & 10.1 & 42 & 3.1 & 0.09 & 12.4 & 0.24 & 3.52 \\
 NGC 4303 & SAB(rs,nl)b\underline{c} & 16.5 & 34 & 7.2 & 0.45 & 18.1 & 0.21 & 3.02 \\
 NGC 4371 & (L)SB$_\mathrm{a}$(r,bl,nr)0$^{0/+}$ & 16.8 & 59 & 3.2 & 0.08 & 15.2 & 0.95 & 5.64 \\
 NGC 4643 & (L)SB(rs,bl,nl)0$^{0/+}$ & 25.7 & 44 & 10.7 & 0.03 & 20.5 & 0.50 & 7.7 \\
 NGC 4981 & SA\underline{B}(s,nl)\underline{b}c & 24.7 & 54 & 2.8 & 0.35 & 13.5 & 0.14 & 2.27 \\
 NGC 4984 & (R´R)SAB$_\mathrm{a}$(l,bl,nl)0/a & 21.3 & 53 & 4.9 & 0.03 & 19.0 & 0.49 & 5.13 \\
 NGC 5236 & SAB(s,nr)c & 7.0 & 21 & 10.9 & 1.95 & 19.3 & 0.37 & 4.02 \\ 
 NGC 5248 & (R´)SAB(s,nr)bc & 16.9 & 41 & 4.7 & 0.40 & 17.3 & 0.49 & 2.41 \\
 NGC 5728 & (R$_1$)SB(\underline{r}´l,bl,nr,nb)0/a & 30.6 & 44 & 7.1 & 0.19 & 20.5 & 0.63 & 9.37 \\ 
 NGC 5850 & (R´)SB(r,bl,nr,nb)\underline{a}b & 23.1 & 39 & 6.0 & 0.11 & 20.8 & 0.80 & 8.20 \\
 NGC 7140 & (R´)SA\underline{B}$_\mathrm{x}$(rs,nrl)a\underline{b} & 37.4 & 51 & 5.1 & 1.29 & 25.8 & 0.63 & 9.25 \\
 NGC 7552 & (R´$_1$)SB(r\underline{s},bl,nr)a & 17.1 & 14 & 3.3 & 0.21 & 21.4 & 0.33 & 4.45 \\
 NGC 7755 & (R´)SAB(rs,nrl)\underline{b}c & 31.5 & 52 & 4.0 & 0.65 & 11.0 & 0.47 & 6.22 \\ 
\hline
 \multicolumn{7}{l}{\textit{\cite{de2023disc}}}  \\
 NGC 289 & (R´,R´L)SAB(rs,rs)\underline{a}b & 18.4 & 43 & 4.0 & -- & 11.0 & -- & 1.96 \\
 NGC 1566 & (R´$_{1}$)SAB(rs,r\underline{s})b & 7.3 & 32 & 3.8 & -- & 9.3 & -- & 1.97 \\
\end{tabular}

\tablebib{(1) \citet{buta2015classical}; (2) NED; (3) \citet{gadotti2019time};
(4) \citet{munoz2015spitzer}; (5) \citet{gadotti2019time}; (6) \citet{munoz2015spitzer}; (7) \citet{gadotti2020kinematic};
(8) \citet{hernquist1993structure}}
\end{table*}

The MUSE-VLT instrument is an integral field spectrograph, with a field of view of $1'\times1'$ and a plate scale of $0.2''/$spaxel, which corresponds to approximately $90,000$ spectra per pointing. The TIMER observations were carried out during Period 97, 2016, from March to October, for $\sim3840~\mathrm{s}$ on average on each source, resulting in a high signal-to-noise ratio (SNR) per pixel (typically above $100$ at the centre). The average PSF FWHM is $0.8''-0.9''$ and the data was reduced using the MUSE pipeline (version 1.6). For more details on the survey and data reduction analysis, we refer the reader to \cite{gadotti2019time}.

\begin{figure}
\centering
\includegraphics[width=\linewidth]{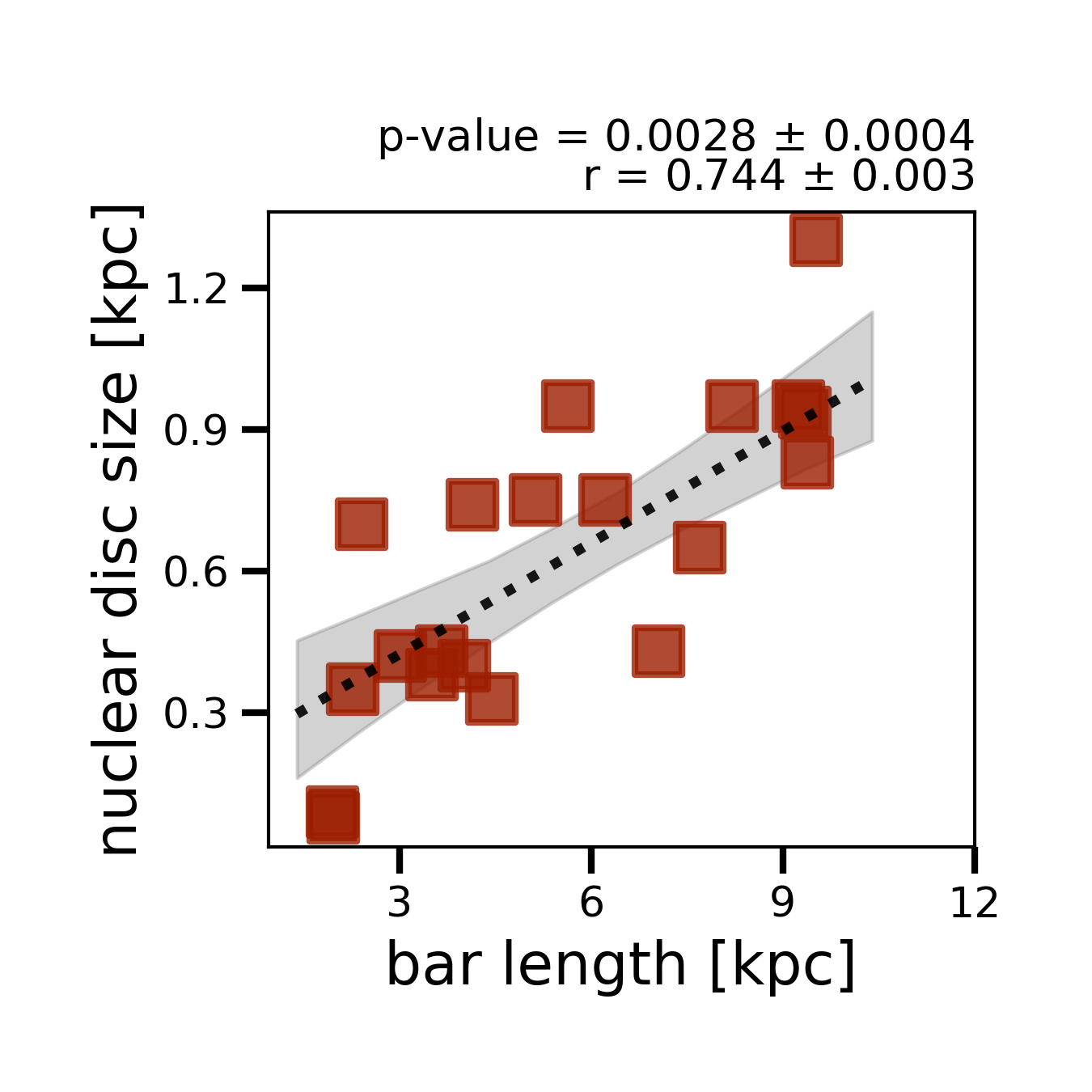}
    \caption{Nuclear disc sizes are visually defined in this work as a function of bar length (see Table~\ref{table_sample}). We bootstrapped our sample $1000$ times to robustly estimate the \textit{Pearson} correlation coefficient and the $p-value$, with associated uncertainties. By different works (e.g., \citealp{gadotti2020kinematic}), we find a strong correlation ($r=0.750\pm0.003$) between the structures, as expected in a bar-built scenario for nuclear discs.}
    \label{fig_rBar_rND}
\end{figure}

For this study, we have selected only those galaxies from TIMER in which a nuclear disc can be identified through a peak in the stellar $\mathrm{V}/\sigma$ radial profile (\citealp{gadotti2020kinematic}). The radius at which this peak occurs, denoted as R$_\mathrm{kin}$, does not necessarily delimit a sharp end of the nuclear disc, but it can be considered a characteristic dynamical radius. This constraint excludes $3$ galaxies from the TIMER sample, which were observed -- these are NGC\,1291, NGC\,1365, and NGC\,6902. Additionally, we have included the galaxies from \cite{de2023disc}, NGC\,289 and NGC\,1566. With that, the final sample considered in this work includes $20$ galaxies (see Table~\ref{table_sample}).  

Even though the R$_\mathrm{kin}$ is dynamically motivated, it is possible to notice star-forming rings placed outside of R$_\mathrm{kin}$ in many cases, which are part of the nuclear disc. Because of that, we visually estimated a new nuclear disc size, R$_\mathrm{ND}$, based on the star formation rate (SFR) and mean stellar age spatial maps. Nevertheless, the correlation between nuclear disc size and bar length discussed in \cite{gadotti2020kinematic} remains strong for the derived nuclear disc sizes in this work (see Figure~\ref{fig_rBar_rND}). 

Lastly, following the analysis in \cite{bittner2020inside}, we split our sample between two sub-samples: low- and high-star-formation (low-SF and high-SF). We measure the median star formation rate surface density ($\Sigma _\mathrm{SFR}$) in the nuclear disc (as defined in this work -- see Table \ref{table_sampleResults}), based on the H$\alpha$ intrinsic luminosity (\citealp{kennicutt1998star}). Although we consider a sharp limit of $5.2\times10^{-7}~\mathrm{M}_\odot\mathrm{yr}^{-1}\mathrm{pc}^{-2}$ to split our sample into two sub-samples with the same size ($10$ galaxies each), we find good agreement with the classification of \cite{bittner2020inside}, in which the authors considered the absolute values and spatial distribution of H$\alpha$ fluxes.

\begin{figure*}
\centering
\includegraphics[width=\linewidth]{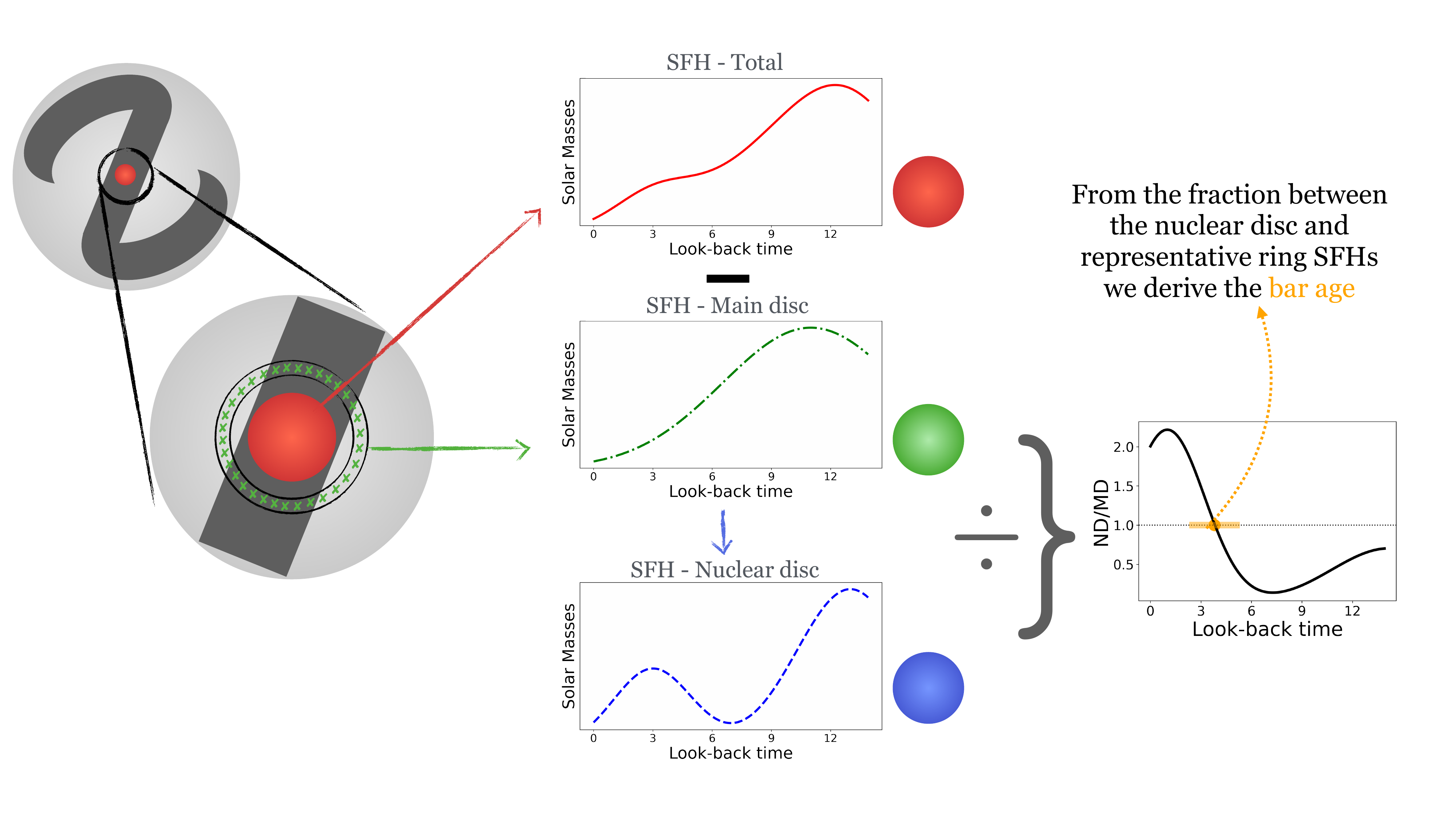}
    \caption{\textbf{Illustration of the bar age measurement method}. In this Figure, we illustrate the method described in Section~\ref{sec_Methodology}, analogous to Figure 2 in \cite{de2023new}. We show the nuclear disc and representative ring regions, from which we derive the SFH of the original data in the nuclear disc region (red) and the representative region (green). We proceed to subtract the main disc SFH from the original one, and the difference is considered the nuclear disc SFH (blue). We then use the ratio ND/MD to time the moment of bar formation (right plot, orange).}
    \label{fig_loom}
\end{figure*}

%##################################################################%
%##################################################################%

\section{Methodology}
\label{sec_Methodology}

As mentioned in the introduction, the ages of the stars in a galaxy's bar are not indicative of its formation epoch, as the bar can form from pre-existing stars in the galaxy's main disc. In \cite{de2023new}, we presented a methodology to separate the populations of the main disc and the nuclear disc, which we briefly describe below.

\subsection{The original method: \cite{de2023new}}

To study the nuclear disc spectra isolated, one needs to remove the contamination of the old stellar populations already present in the region in which the nuclear disc was formed. In \cite{de2023new}, we developed a methodology to separate the spectra from the nuclear disc and the main disc. Concisely, the method is based on the assumptions that (1) the populations of the main disc do not change significantly in the regions where the nuclear disc resides, and (2) the brightness of the main disc varies with radius following an exponential law. 

We use MUSE datacubes and consider the spectrum of a ring region around the nuclear disc as a proxy for the stellar population of the main disc -- which was already present before the formation of the bar. We use this spectrum and the disc scale length of the galaxy (measured in \citealp{salo2015spitzer}) to model an exponential main disc in the nuclear disc region and subtract it from the observed data. To perform this subtraction between spectra and to avoid creating spectral line artefacts, we shift all the spectra to the rest frame and convolve the lines to the same velocity dispersion, considering the kinematic results from a previous analysis. Additionally, we use the BPT diagram diagnosis (\citealp{baldwin1981classification}) to avoid AGN-dominated spaxels. Once we remove the modelled main disc from the data, we employ the GIST pipeline (\citealp{bittner2019gist}) to derive independent star formation histories (SFHs) on the nuclear and main discs.

For the GIST run, we followed the configuration from previous TIMER papers to guarantee consistency between analyses (e.g., \citealp{gadotti2019time}; \citealp{de2019clocking}; \citealp{bittner2020inside}; \citealp{de2023new}; \citealp{de2023disc}). We considered a wavelength range of $4800-5800~\AA$ and Voronoi-binned (\citealp{cappellari2003adaptive}) the data to a final signal-to-noise ratio of 100. Additionally, we used MILES simple stellar population libraries (SSPs -- \citealp{vazdekis2015evolutionary}), with values of [M/H] varying between $-1$ and $0.04$, ages between $0.03$ and $14~\mathrm{Gyr}$, and $\alpha$-enhancements of $0$ and $+0.4$ -- we would like to highlight that for this wavelength range, $\alpha$ measurements are dominated by magnesium (Mg). Firstly, GIST employs an unregularized pPXF (\citealp{cappellari2004parametric}; \citealp{cappellari2017improving}) run to derive stellar kinematics. We also include a low-order multiplicative Legendre polynomial to account for mismatches between the observed spectra and the continuum templates. From this step, we derive the stellar velocity, stellar velocity dispersion, $h_3$, and $h_4$. Secondly, GIST employs pyGandALF (\citealp{sarzi2006sauron}; \citealp{falcon2006sauron}) to model and subtract emission lines as Gaussians, resulting in the emission-subtracted spectra. Lastly, GIST employs a regularized pPXF run to the emission-subtracted spectra, fixing the kinematic information from the first step, and fitting different templates of the stellar population from MILES. From the last step, we retrieve the spatial distribution of light-weighted mean stellar population properties -- age, [M/H], and [$\alpha$/Fe]. In addition, the regularized pPXF run results in the weight of each SSP. That is the fraction of the light that each SSP accounts for -- for different ages, metallicities, and $\alpha$-enhancements. Considering the weights for a given age bin and mass-to-light fractions\footnote{http://research.iac.es/proyecto/miles/pages/predicted-masses-and-photometric-observables-based-on-photometric-libraries.php} (\citealp{vazdekis2015evolutionary}), we can reconstruct the mass-weighted SFH for the nuclear disc and the main disc, independently. We consider the independent SFHs to measure the moment of bar formation.

To test the validity of the described methodology, we reproduced in \cite{de2023new} the same approach in simulated galaxies, for which we know the moment the bar has formed. With this test, we also aimed to understand limitations and possible contamination in the nuclear disc clean spectra -- and consequently, SFHs. We concluded that a good proxy for timing bar formation is the moment in which the SFH of the nuclear disc surpasses the SFH of the main disc ($\mathrm{ND} / \mathrm{MD} > 1$) with a positive slope towards younger ages (i.e., when the SFR in the nuclear disc is higher than in the main disc). We would like to stress that this criterion can be a lower limit in some cases. In the scenario in which the SFR of the main disc is still ongoing in the centre when the bar forms, and the gas inflow due to the bar is low -- consequently, the SFR in the nuclear disc as well -- the criterium of $\mathrm{ND} / \mathrm{MD} > 1$ might not be sufficient at the moment of bar formation and the bar age we derive would be a lower limit. Nevertheless, we do not expect this to be the case in many galaxies since if the main disc is still forming stars in the centre, the galaxy is most likely gas-rich, and the gas inflow due to the bar is unlikely to be significantly lower. Additionally, studies find that galaxies quench inside-out (e.g., \citealp{ellison2018star}; \citealp{belfiore2018sdss}), so it is expected that the central region of the main disc will present lower SFR than the outer region.\color{black} We refer the reader to \cite{de2023new} for further details and tests.

\subsection{This work: complementary approach}

The methodology described above was employed in \cite{de2023new} for one galaxy and in \cite{de2023disc} for two galaxies. However, it demands considerable computational time, especially in shifting and convolving the spaxel-by-spaxel spectra. Since our goal is to apply said methodology to a larger sample, we tested a complementary approach to derive bar ages, keeping the same strategy: we use a ring region to model the main disc and subtract it from the original data. The difference lies in the fact that instead of disentangling spectra of different structures -- which demands convolving, shifting, and normalizing (further details in \citealp{de2023new}) --, we work with the SFHs. We derive the SFHs spaxel-by-spaxel for the original data cube and proceed with disentangling the nuclear disc from the main underlying disc directly on the SFHs, instead of applying this process to the spectra (see Figure~\ref{fig_loom}). In this work, we derived results from both approaches  -- spectra vs. star formation histories -- for most galaxies, finding agreeable results with differences within $1.0~\mathrm{Gyr}$, which is comparable to the uncertainties of the original method. The original approach took considerably longer to perform and, since we aim to apply this methodology to larger samples in the near future, we consider a good compromise to use the new approach with larger systematic errors. However, we were not able to derive statistical errors for this sample, which could vary depending on case by case -- especially in the case of older bars, in which older stellar populations have higher uncertainties. Considering the systematic errors from \cite{de2023new} and the new approach used in this work, we estimate the final average systematic errors of $^{+1.8}_{-1.4}~\mathrm{Gyr}$ -- nonetheless, note that older bars have intrinsic higher uncertainties associated to older stellar populations.

\begin{figure}
\centering
\includegraphics[width=\linewidth]{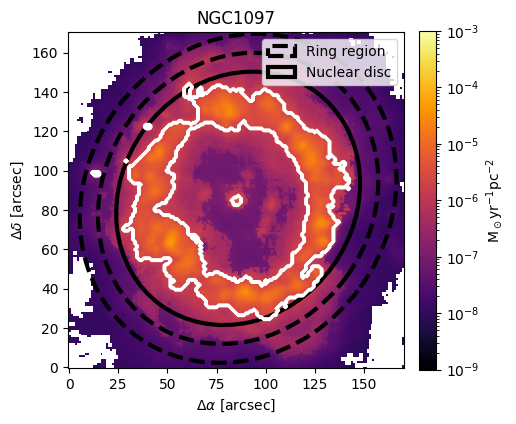}
    \caption{Star formation rate surface density map of NGC\,1097 ($\Sigma_\mathrm{SFR}$ in units of M$_\odot~yr^{-1}~pc^{-2}$). We present the limits of the nuclear disc defined in this work (Table \ref{table_sampleResults} -- solid-black contour) and the ring region from which we extract the representative spectrum in dashed-black contour -- see illustration in Fig. \ref{fig_loom}. Additionally, we present the $\Sigma_\mathrm{SFR} = 2\times10^{6}$ mask (solid-white countour). As one can see, the mask is limited to the star-forming ring. Since we are interested in the oldest stars that belong to the nuclear disc, masking the active star-forming regions will not affect our methodology.}
    \label{fig_dSFRNGC1097}
\end{figure}

Additionally, considering that some of the galaxies in our sample have ongoing star formation in the nuclear disc region, we decided to mask spaxels with $\Sigma_\mathrm{SFR}~\geq~ 2\times10^{6}~\mathrm{M}_\odot \mathrm{yr}^{-1}\mathrm{pc}^{-2}$, a typical value we find in star-forming rings in our sample (see for example NGC\,1097 in Figure~\ref{fig_dSFRNGC1097}). The only exception is NGC\,7552, for which all spaxels in the nuclear disc region have high star formation, and thus, in this case, we masked spaxels with $\Sigma_\mathrm{SFR}~\geq~ 5\times10^{6}~\mathrm{M}_\odot \mathrm{yr}^{-1}\mathrm{pc}^{-2}$, instead. This is because MILES libraries are not well suited for very young stellar populations, and their light can outshine older stellar populations, which is our interest. Nevertheless, we retrieve results both for masking and not masking the star-forming regions, finding differences in the bar age estimates within $0.9~\mathrm{Gyr}$. 

\subsection{Multiple nuclear components}
\label{subsubsec_MultipleNuclearComponents}

Although no large classical bulges with high stellar velocity dispersion have been found in the TIMER sample (\citealp{bittner2020inside}), other structures may be present in the centre of the galaxies, co-existing with the nuclear disc. As demonstrated by \cite{mendez2014secular}, some galaxies can have a small classical bulge, embedded inside the nuclear disc. In this case, the centre of the nuclear disc shows a dispersion-dominated, separate structure. This second structure, if present in our sample, is not accounted for by our methodology when extrapolating the exponential profile of the main disc towards the central region. Furthermore, depending on the formation scenario of this small classical bulge, the stellar populations and SFH may be completely different as compared to the nuclear disc. In this case, there could be contamination on the nuclear disc's ``clean'' SFH. On the other hand, substructures of the nuclear disc itself, such as inner/nuclear bars and nuclear spirals, can also increase the velocity-dispersion, although they are expected to be part of the evolution history of the nuclear disc. Due to that, one cannot simply mask regions of higher velocity-dispersion assuming it to be an independent structure, and a detailed stellar population diagnosis should be performed. 

\renewcommand{\arraystretch}{1.5}
\begin{table}
\caption{List of objects with multiple nuclear components, with the final ages considered are in boldface.}
\centering
\begin{tabular}{lccc}
 \textbf{Galaxy} & \textbf{Nuclear bar}  & \textbf{Bar Age (masked)} & \textbf{Bar age} \\
 \hline \hline
IC\,1438 & Y  & -- & \textbf{10.0} \\
NGC\,1097 & N & \textbf{4.0} & $5.5$ \\
NGC\,1433 & Y & -- & \textbf{9.5} \\
NGC\,4371 & N & \textbf{10.5} & -- \\
NGC\,4643 & N & \textbf{13.0} & -- \\
NGC\,4981 & N & \textbf{3.75} & -- \\
NGC\,4984 & N & \textbf{3.0} & $5.5$ \\
NGC\,5728 & Y & -- & \textbf{7.5} \\
NGC\,5850 & Y & -- & \textbf{13.5}
\end{tabular}
\label{table_MultipleNuclearComponents}
\end{table}

In our sample of $20$ galaxies, $9$ displayed an increase in the velocity dispersion in the central region; they are IC\,1438, NGC\,1097, NGC\,1433, NGC\,4371, NGC\,4643, NGC\,4981, NGC\,4984, NGC\,5728, and NGC\,5850 (see Table~\ref{table_MultipleNuclearComponents}). Since constructing a detailed diagnosis of substructures is beyond the scope of this work, we decided whether to mask or not these regions, case by case. Amongst these nine galaxies, four have inner/nuclear bars reported; these are IC\,1438 (TIMER collaboration, in prep.), NGC\,1433 (\citealp{erwin2004double}; \citealp{buta2015classical}; \citealp{bittner2021galaxies}), NGC\,5728 (e.g., \citealp{erwin2004double} -- although the presence of the nuclear/inner bar in this case is ambiguous, see \citealp{de2019clocking}), and NGC\,5850 (\citealp{de2013distinct}). We did not mask the central region of these four galaxies, since the inner/nuclear bar is expected to be part of the nuclear disc evolution and share the same SFH. For the rest, we derive two bar ages: masking the central high-velocity dispersion region [denoted in Table~\ref{table_MultipleNuclearComponents} as `bar age (masked)'] and not masking it (denoted as `bar age'), listed in Table~\ref{table_MultipleNuclearComponents}. For NGC\,4371, NGC\,4643, and NGC\,4981, we cannot derive bar ages in the configuration where we do not mask the central region, since the criterium of $\mathrm{ND} / \mathrm{MD} > 1$ is not fulfilled (see Section \ref{sec_Methodology}). For the remaining two galaxies, NGC\,1097 and NGC\,4984, the change in bar age is $1.5$ and $2.5~\mathrm{Gyr}$, respectively, where the non-masked results tend to be older. This can be due to two scenarios: (\textit{i}) the nature of the central velocity dispersion reflects indeed a small classical bulge, which is older and biases our results towards older ages; or (\textit{ii}) the nuclear disc grows inside-out and the central region formed first (\citealp{bittner2020inside}), so we would be retrieving a lower limit for the bar age by masking this region. Since it is not yet conclusive whether the nuclear disc actually reaches the very centre of the galaxy or has an inner edge, and in the presence of the high-velocity dispersion, we decided to consider the values when masking the central region. In summary, four galaxies have inner/nuclear bars, so we did not mask the very central region with high-velocity dispersion (IC\,1438, NGC\,1433, NGC\,5728, and NGC\,5850), and five galaxies have bar ages with the very central region masked (NGC\,1097, NGC\,4371, NGC\,4643, NGC\,4981, and NGC\,4984). 

%##################################################################%
%##################################################################%

\section{Results}
\label{sec_results}

We applied the methodology described in Sect.~\ref{sec_Methodology} to all TIMER galaxies with a discernible nuclear disc, listed in Table~\ref{table_sample}. We successfully derived bar ages for $18$ galaxies and added the results from \cite{de2023disc}, resulting in the largest sample of known bar ages, with a total of $20$ galaxies (Table~\ref{table_sampleResults}). In this Section, we present our main results and the first insights on secular evolution derived from our age-dating of the bars.

\label{subsubsec_barAges}

\renewcommand{\arraystretch}{1.5}
\begin{table*}[!h]
\caption{List of derived bar ages, nuclear disc sizes and masses.}
\label{table_sampleResults}
\centering
\begin{tabular}{lccc|lccc}
\textbf{Galaxy} & \textbf{Bar age} & \textbf{R$_{\mathrm{ND}}$} & \textcolor{black}{\textbf{ND mass}}  & \textbf{Galaxy} & \textbf{Bar age} & \textbf{R$_{\mathrm{ND}}$} & \textcolor{black}{\textbf{ND mass} }\\ 
& $^{+1.4}_{-1.8}$ [Gyr] & [Kpc] & \textcolor{black}{[$10^{9}$ M$_\odot$]} & & $^{+1.4}_{-1.8}$ [Gyr] & [Kpc] & \textcolor{black}{[$10^{9}$ M$_\odot$]} \\
 \hline \hline
\multicolumn{3}{l}{\textit{Low-star-formation}} & & \multicolumn{3}{l}{\textit{High-star-formation}}  \\
IC1438 & $11.5$ & 0.740 & \textcolor{black}{2.28} & NGC613 & $2.5$ & 0.830 & \textcolor{black}{1.07} \\
NGC1300 & $10.5$ & 0.430 & \textcolor{black}{0.64} & NGC1097 & $4.0$ & 1.300 & \textcolor{black}{1.9} \\
NGC1433 & $9.5$ & 0.430 & \textcolor{black}{0.59} & NGC3351 & $6.5$ & 0.380 & \textcolor{black}{0.31} \\ 
NGC4371 & $11.0$ & 0.950 & \textcolor{black}{1.43} & NGC4303 & $6.0$ & 0.420 & \textcolor{black}{0.67} \\ 
NGC4643 & $13.0$ & 0.650 & \textcolor{black}{5.31} & NGC4981 & $3.75$ & 0.350& \textcolor{black}{0.24}   \\
NGC4984 & $3.0$ & 0.750 & \textcolor{black}{0.25} & NGC5236 & $2.25$ & 0.400 & \textcolor{black}{0.11} \\
NGC5248 & $3.5$ & 0.700 & \textcolor{black}{0.49} & NGC5728 & $7.5$ & 0.935 & \textcolor{black}{1.13} \\
NGC5850 & $13.5$ & 0.950 & \textcolor{black}{3.39} & NGC7552 & $2.75$ & 0.330 & \textcolor{black}{0.09} \\ \cline{5-8}
NGC7140 & $6.0$ & 0.950 & \textcolor{black}{0.44} & \multicolumn{3}{l}{\textit{From \cite{de2023disc}}}   \\
 NGC7755 & $10.5$ & 0.750 & \textcolor{black}{1.13} & NGC289 & $4.5$ & 0.090 & \textcolor{black}{0.04} \\
 & & & & NGC1566 & $0.7$ & 0.077 & \textcolor{black}{0.03} \\
\end{tabular}
\end{table*}

\begin{figure*}
\centering
\includegraphics[width=0.7\linewidth]{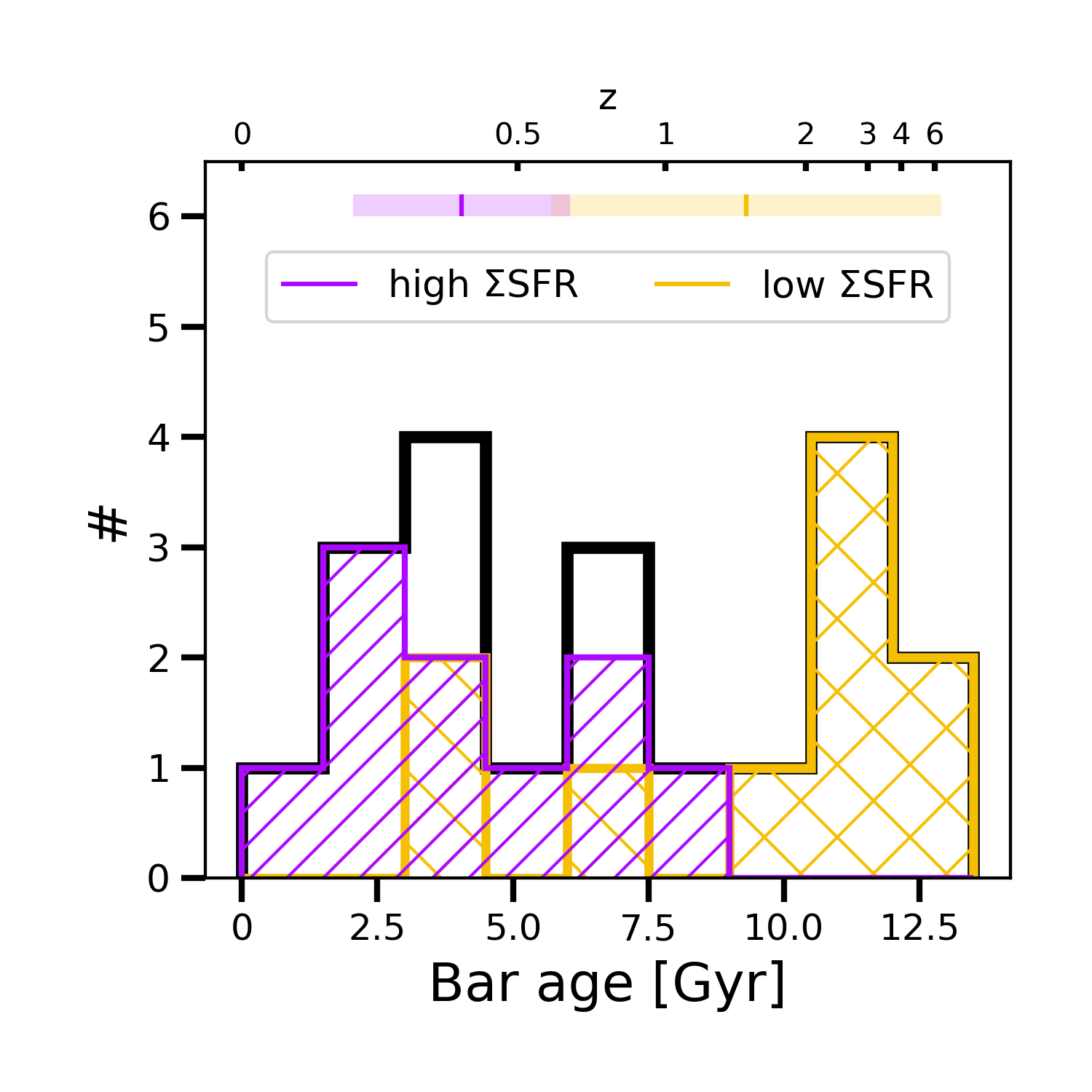}
    \caption{\textbf{Distribution of bar ages derived in this work}. We display our main results (total in black line), colour-coded by low- and high-star-formation sub-samples (yellow and purple, respectively). On the top, we present the median bar age of each sub-sample, together with the standard deviation of each distribution (high-SF: $4.0\pm2.0~\mathrm{Gyr}$; low-SF: $9.3\pm3.6~\mathrm{Gyr}$). It is clear that low-SF nuclear discs are hosted by older bars (typically older than $9~\mathrm{Gyr}$), whereas high-SF nuclear discs are hosted by younger bars. Additionally, we derive a large range of bar ages, varying from $1-13~\mathrm{Gyr}$, illustrating that bars started to form in a young Universe ($\sim1~\mathrm{Gyr}$), and this is an ongoing process in the Local Universe.}
    \label{fig_histBarAges}
\end{figure*}

\begin{figure}
\centering
\includegraphics[width=\linewidth]{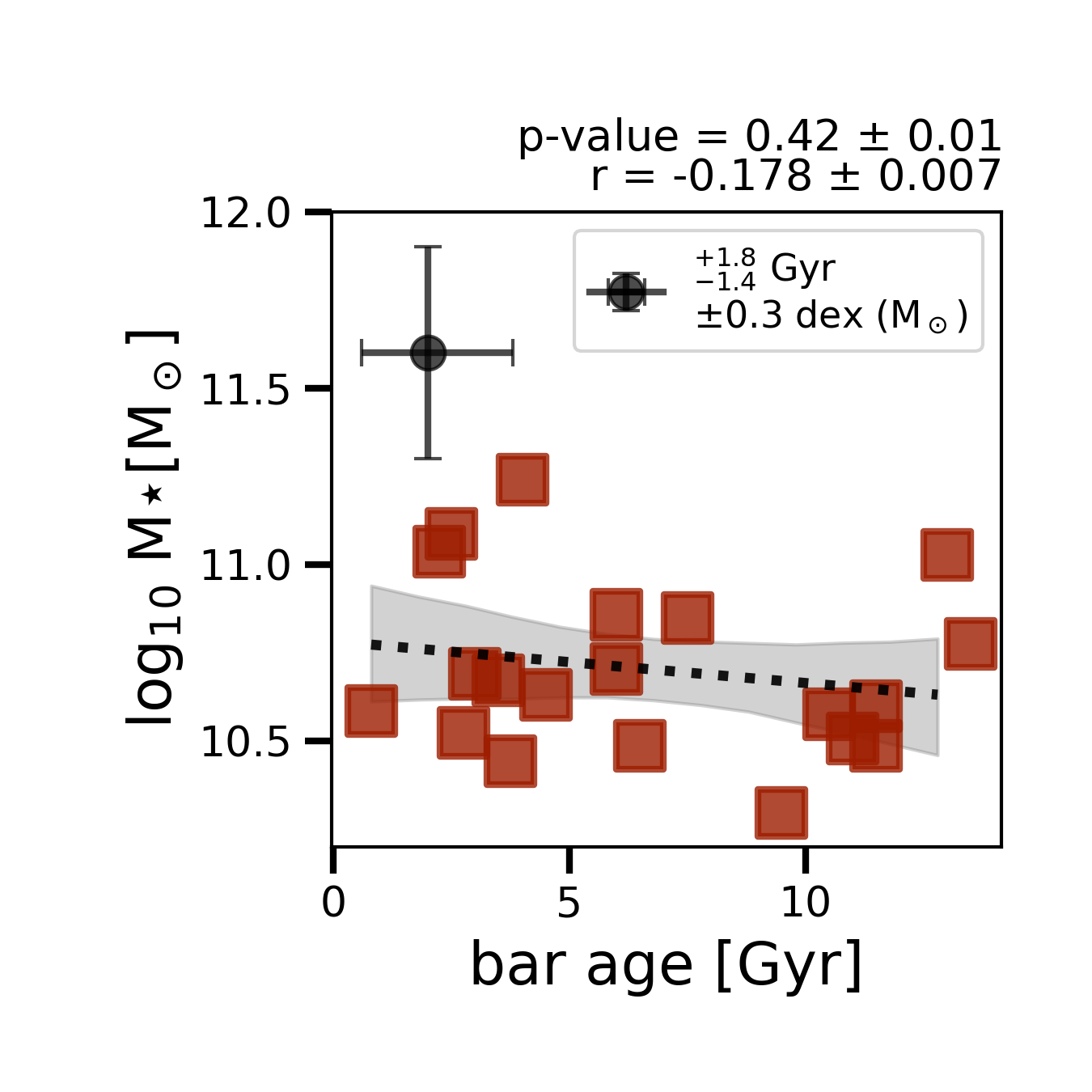}
    \caption{\textbf{Stellar masses as a function of bar ages}. We consider the stellar masses from the S$^4$G survey, obtained with the $3.6\mu\mathrm{m}$ band (\citealp{munoz2015spitzer}) and the bar ages derived in this work.  We bootstrapped our sample $1000$ times to robustly estimate the \textit{Pearson} correlation coefficient and the $p-value$, with associated uncertainties. Lastly, we display the mean errors in the bar age \color{black} and stellar mass \color{black}estimates (black-dot). \color{black} As discussed in \cite{munoz2015spitzer}, the main source of error in the galaxy stellar mass comes from errors in distance estimates, which translate to $\pm0.32$ dex error in mass. \color{black} Contrary to downsizing predictions, we find no correlation between the two properties, with a weak Pearson coefficient of $r=-0.180\pm0.007$ -- illustrated by the dotted line.}
    \label{fig_barAge_mstar}
\end{figure}

% \subsection{Secular evolution}

\begin{figure*}[!ht]
\centering
\includegraphics[width=0.9\linewidth]{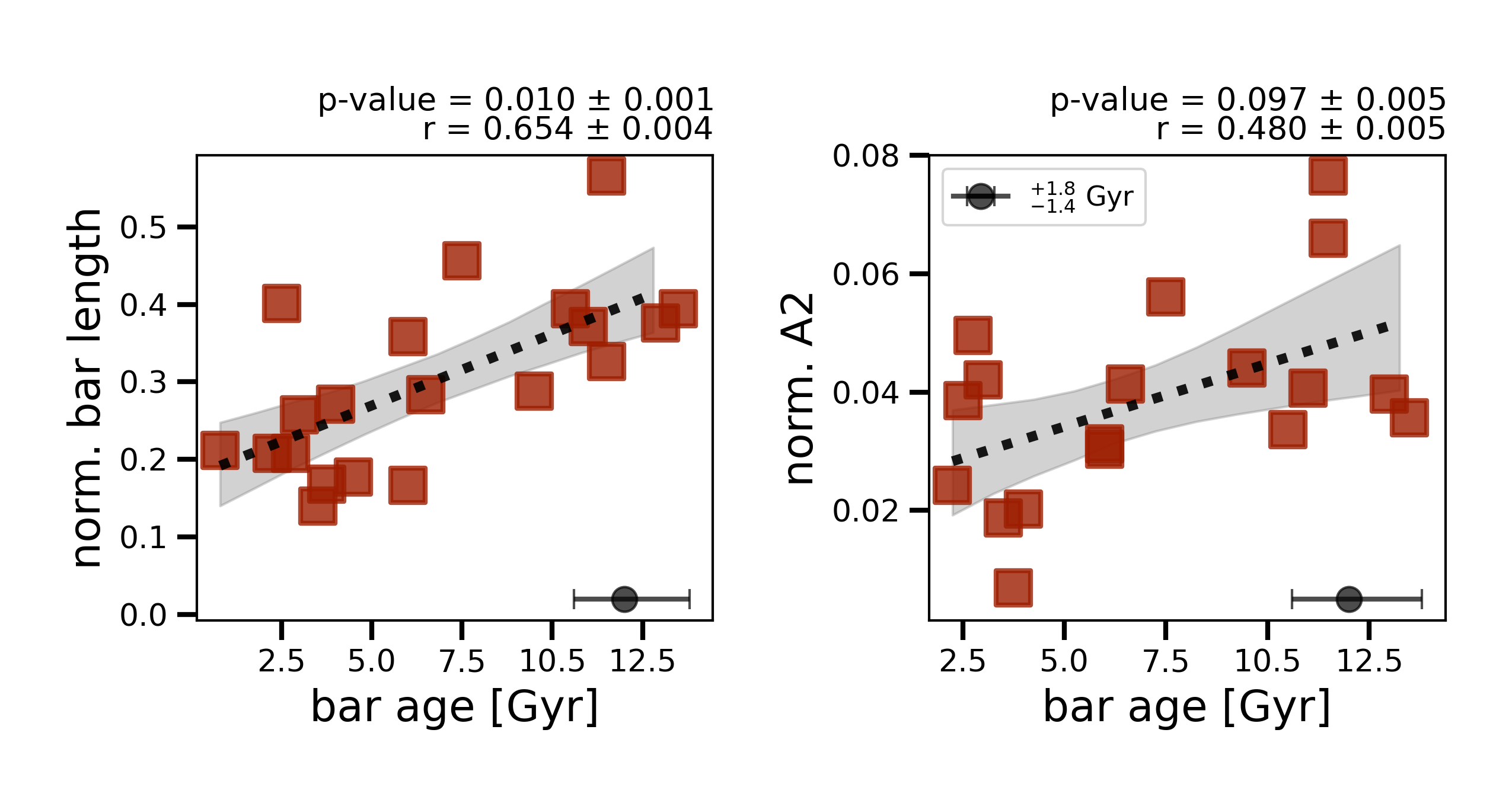}
    \caption{\textbf{Insights on bar ageing.} We display the normalized bar lengths (left panel) and normalized bar strengths (A2- right panel) for our sample by galaxy size. We also present the linear regression (dotted-black line) for $1000$ bootstrap repetitions, the Pearson correlation coefficient $r$, and the associated null hypothesis $p-value$. For the two properties, we find trends of correlation. For the bar length, the associated $p-value$ is smaller than $0.05$, while for A2, its $p-value\sim1$. These results are in agreement with the scenario in which older bars are longer and stronger.}
    \label{fig_barAge_nD_bar_A2}
\end{figure*}

\begin{figure}
\centering
\includegraphics[width=\linewidth]{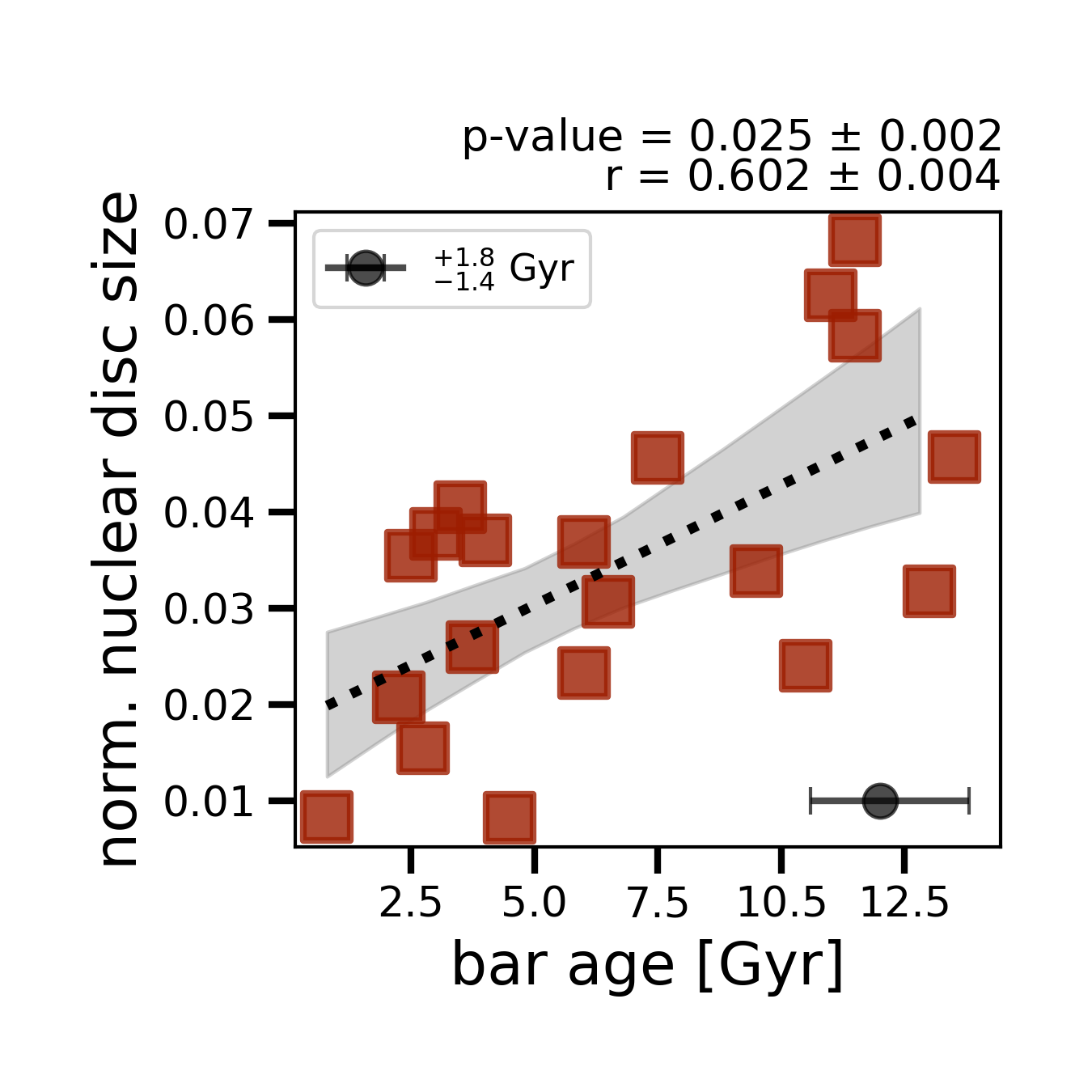}
    \caption{\textbf{Nuclear disc size normalized by galaxy size with respect to bar ages}. We present the nuclear disc size normalized by the galaxy size (see Tables~\ref{table_sample} and \ref{table_sampleResults}) for different bar ages. Additionally, we display the linear regression for $1000$ bootstrap repetitions, with the Pearson coefficient value of $r=0.600\pm0.004$ and $p-value=0.029\pm0.003$. From this result, it is clear that older bars tend to host larger nuclear discs, while younger bars have smaller ones. This is consistent with the scenario in which nuclear discs grow with time, following the inside-out evolution context proposed by \cite{bittner2020inside}.}
    \label{fig_barAge_ND}
\end{figure}

\begin{figure}
\centering
\includegraphics[width=\linewidth]{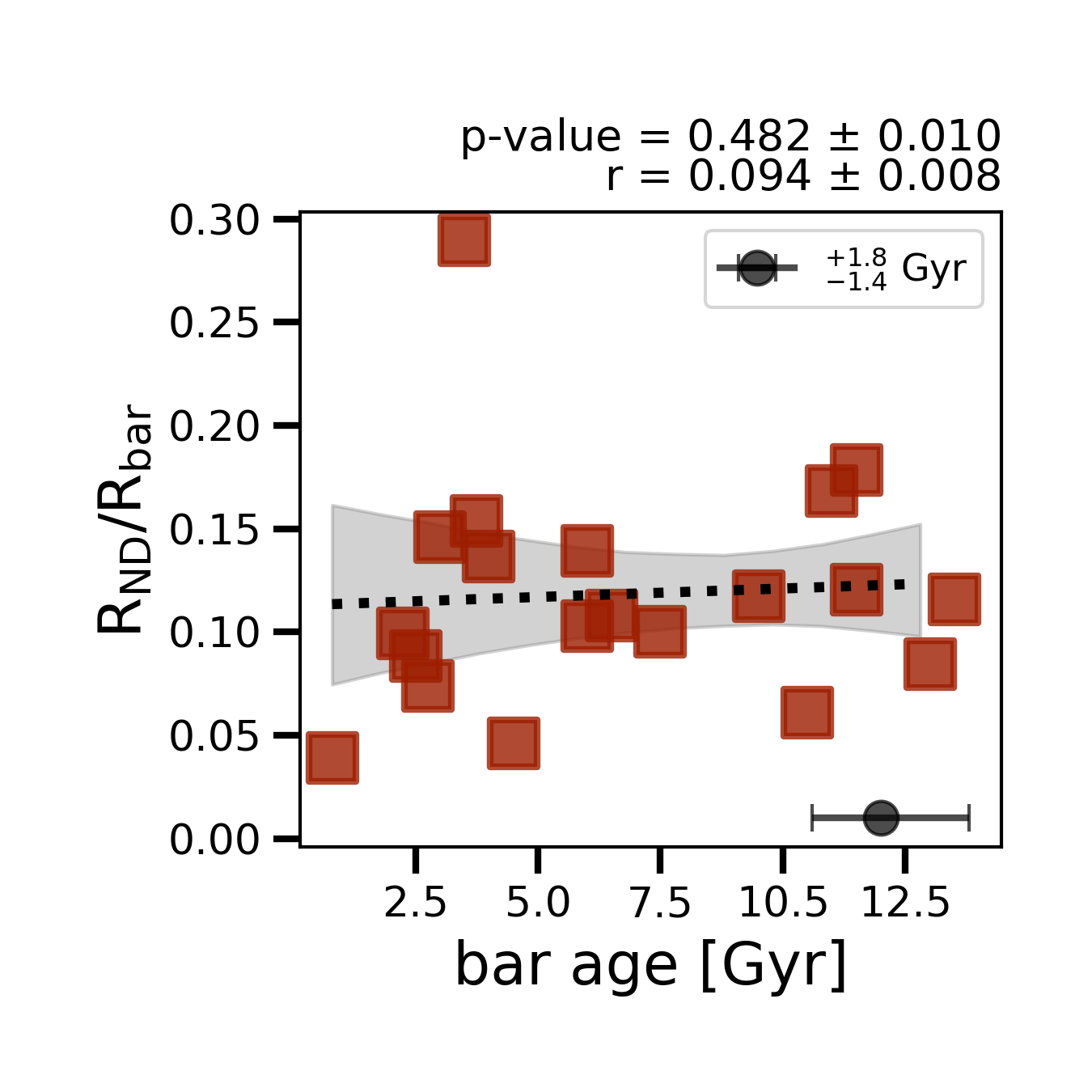}
    \caption{\textbf{Nuclear disc normalized by bar length with respect to bar ages}. We present the nuclear disc size normalized by the bar length (see Tables \ref{table_sample} and \ref{table_sampleResults}) for different bar ages. It is clear that the size relation between the nuclear disc and the bar does not depend on the bar age, with a constant value close to $\sim12\%$. This demonstrates how the nuclear disc sizes depend on bar properties and resonances, and as the bar evolves, the nuclear disc's absolute size evolves as well.}
    \label{fig_barAge_NDRbar}
\end{figure}

\begin{figure}
\centering
\includegraphics[width=\linewidth]{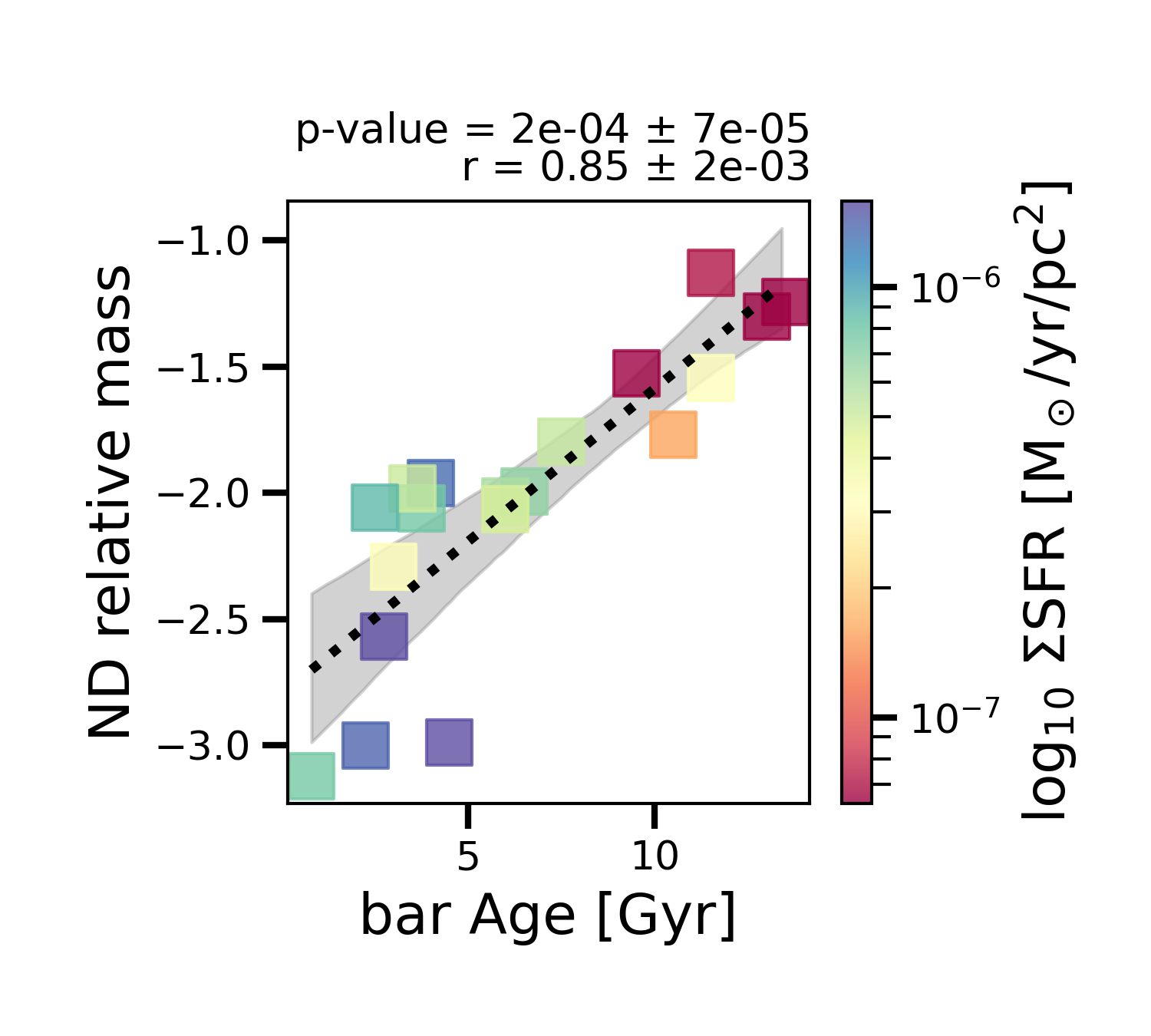}
    \caption{\textcolor{black}{\textbf{Nuclear disc mass build-up:} In this plot, we present the nuclear disc mass normalized by the galaxy mass for different bar ages. To measure the nuclear disc mass, we integrated the isolated SFH (blue-dashed curves in Figs.~\ref{fig_lSFmain} and \ref{fig_hSFmain}), following the same approach in \cite{de2023new,de2023disc}. Additionally, we color-code each point accordingly to its $\Sigma_\mathrm{SFR}$, which increases towards blue. It is clear that young nuclear discs are less massive with a higher star formation rate. As the bars age, the nuclear discs gradually present less star formation, while building up their mass. Lastly, we find a correlation coefficient of 0.85$\pm2\times10^{-3}$, with $p-value=2\times10^{-4}\pm7\times10^{-5}$.}}
    \label{fig_barAge_NDmass}
\end{figure}

\begin{figure*}
\centering
\includegraphics[width=1\linewidth]{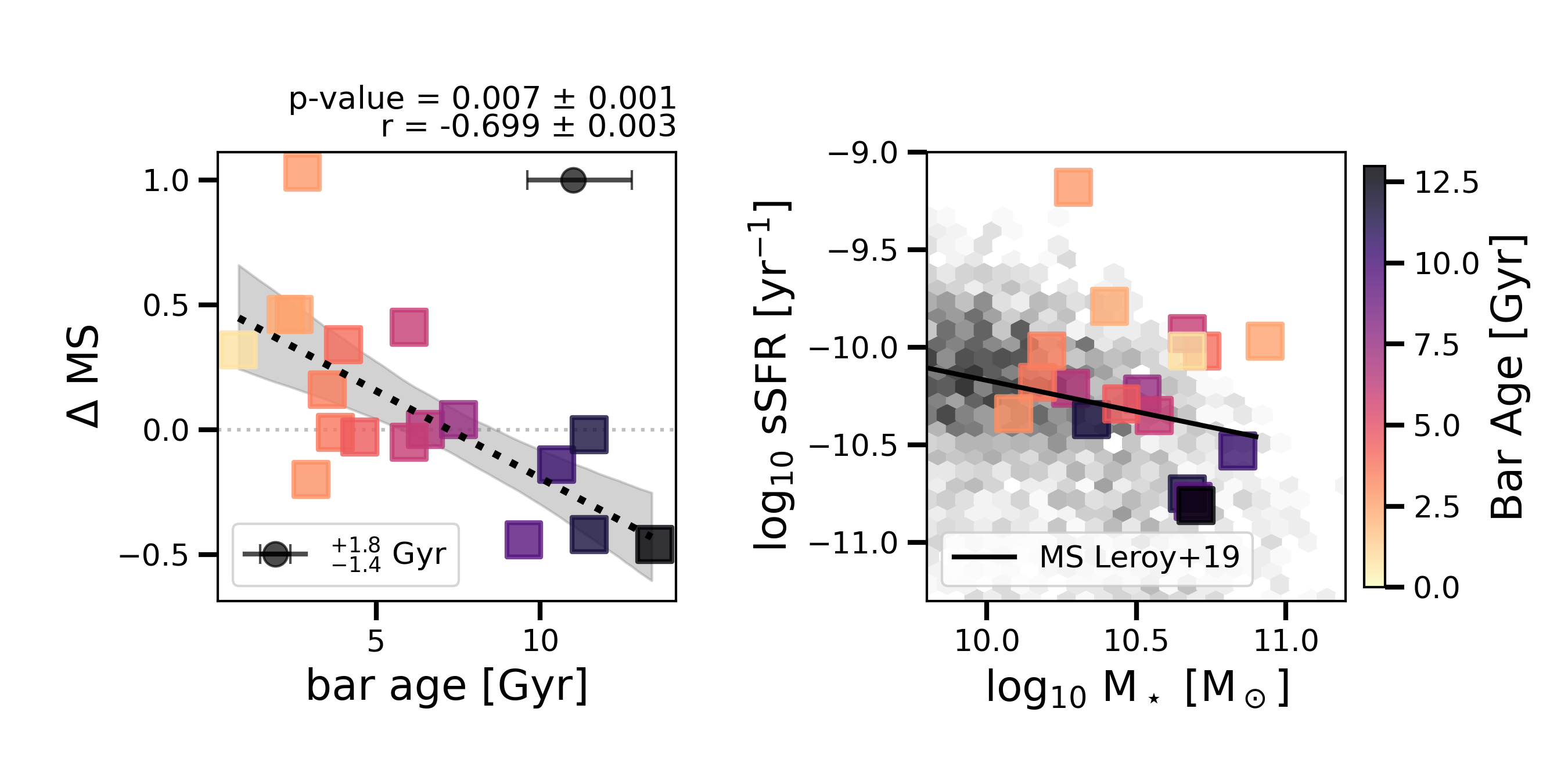}
    \caption{\textbf{Galaxy star formation properties with respect to bar ages}. In this Figure, we investigate how the bar ageing is related to the star formation of the host galaxy. \textbf{Left: } We display the main sequence offset ($\Delta\mathrm{MS}$ -- values from \citealp{leroy2019z}) of the host galaxy with respect to the bar age. Positive values of $\Delta\mathrm{MS}$ indicate the galaxy is bursting, while negative, quench. Additionally, we bootstrap our sample $1000$ times, in the gray-shaded area, finding a correlation coefficient of $\mathrm{r}=-0.699\pm0.003$, with $p-value=0.007\pm0.001$. The strong correlation between star formation of the host galaxy and bar ageing is in agreement with the scenario in which bars aid the quenching of the galaxy. \textbf{Right: } we display the main sequence of galaxies defined in \cite{leroy2019z}, that is, the sSFR with respect to the galaxy mass. We show their entire sample in a grey density map while highlighting our sample colour-coded according to bar age, contextualizing our results.}
    \label{fig_barAge_quench}
\end{figure*}

\begin{figure}
\centering
\includegraphics[width=\linewidth]{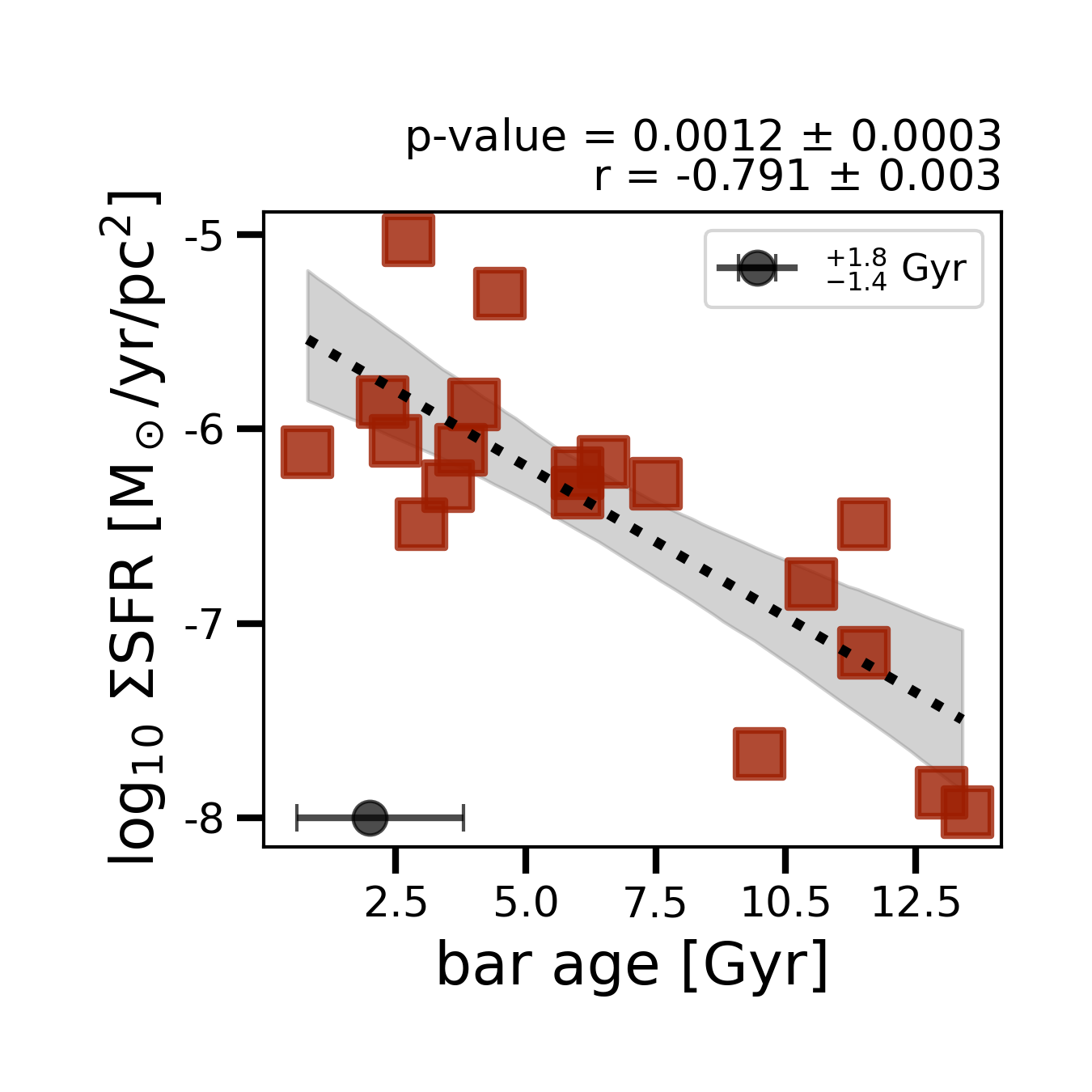}
    \caption{\textbf{$\Sigma_\mathrm{SFR}$ in the nuclear disc with respect to bar ageing}. We present the measured median $\Sigma_\mathrm{SFR}$ in the nuclear disc region (as defined in this work -- see Table \ref{table_sampleResults})
    for different bar ages, finding a strong anti-correlation ($r=-0.791\pm0.003$ and $p-value=0.0011\pm0.0003$). This demonstrates that, as the bar ages, the nuclear discs tend to form fewer stars.}
    \label{fig_barAge_dSFR}
\end{figure}

\begin{figure}
\centering
\includegraphics[width=\linewidth]{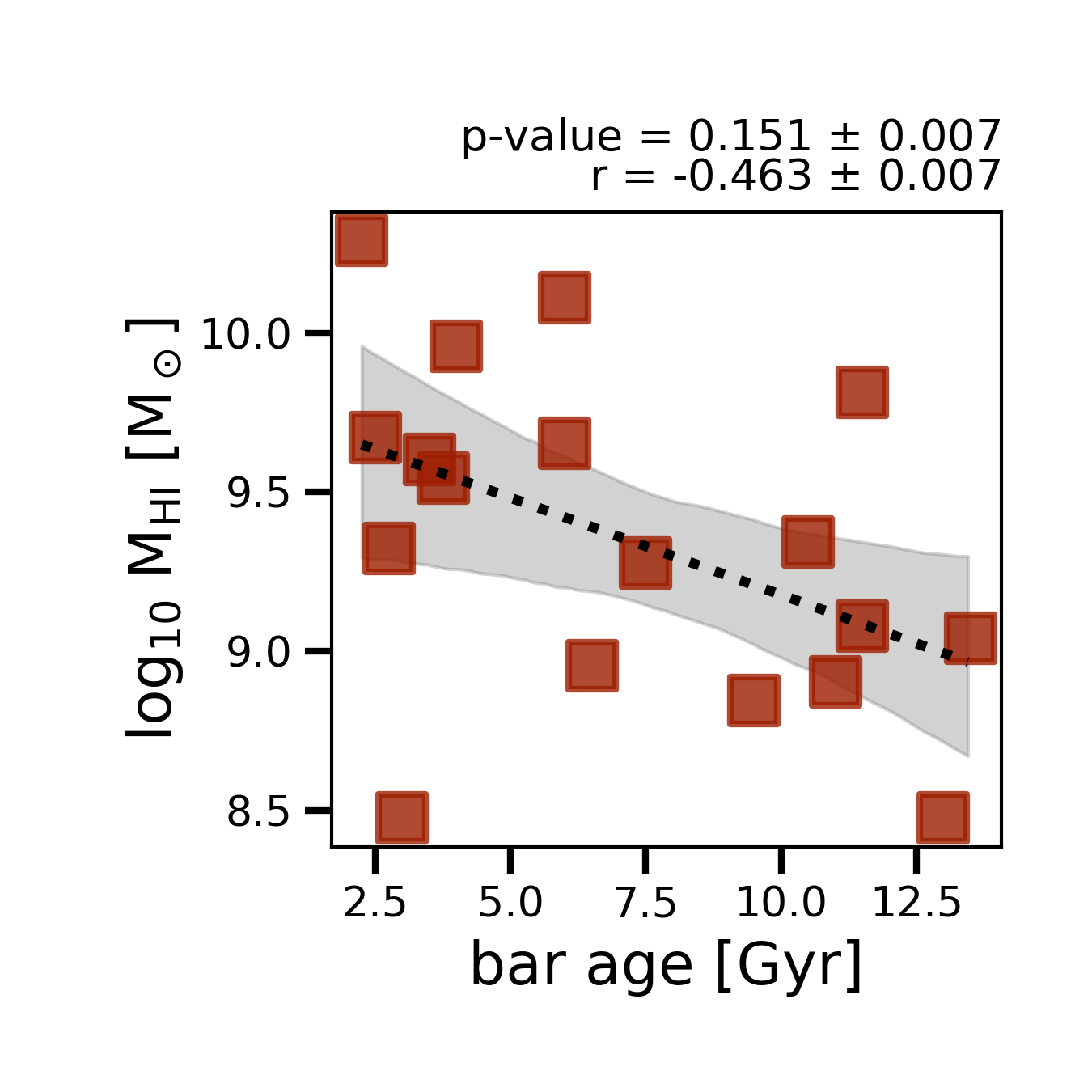}
    \caption{\textbf{HI mass of the host galaxy with respect to bar ages}. We present the mass of neutral hydrogen from \cite{gadotti2019time}, based on the 21-cm line fluxes available from LEDA (see Table \ref{table_sample}), finding a weak trend between HI mass and bar age, with $r=0.463\pm0.007$, which is not statistically significant, with $p-value=0.151\pm0.007$. This weaker anti-correlation indicates that even though the bar can aid in the quenching of the host galaxy, the gas of the host galaxy is not being completely exhausted, but star formation is rather less efficient (e.g., \citealp{saintonge2016molecular}; \citealp{bacchini2019volumetric}; \citealp{pessa2022variations}).}
    \label{fig_barAge_mHI}
\end{figure}

We find a wide range of bar formation epochs that vary between mass-weighted ages of $1.0-13.0~\mathrm{Gyr}$ (see Fig.~\ref{fig_histBarAges}). The high-SF nuclear discs sub-sample (see Sec .~\ref{sec_SampleDataDescr}) tends to have bars younger than $9~\mathrm{Gyr}$, whereas the low-SF nuclear disc sub-sample hosts older bars, the majority of them with ages greater than $9~\mathrm{Gyr}$. The mean bar ages of high- and low-SF nuclear discs are $4.0 \pm 2.0~\mathrm{Gyr}$ and $9.3 \pm 3.6~\mathrm{Gyr}$, respectively. The result is not driven by the spaxels with large SFR values, since we mask the ones with values above $2\times10^{6}~\mathrm{M}_\odot \mathrm{yr}^{-1}\mathrm{pc}^{-2}$. Three galaxies from the low-SF sub-sample present bar ages of $3.0$, $3.5$, and $6.0~\mathrm{Gyr}$; they are NGC\,4984, NGC\,5248, and NGC\,7140, respectively, where the first had the central region masked (see Section.~\ref{subsubsec_MultipleNuclearComponents}) and the last two have $\Sigma_\mathrm{SFR}$ close to the sharp limit that defines the sub-samples, which can explain the relatively young ages compared to the rest of the low-SF sub-sample. We show the star formation histories and bar ages derived for each galaxy in the appendix in Figures~\ref{fig_lSFmain} and~\ref{fig_hSFmain}. 

In the downsizing scenario of bar formation (see \citealt{sheth2012hot}), it is expected that the most massive galaxies would have achieved the necessary mass to become self-gravitating first, which would suggest that older bars would be hosted by the most massive galaxies. To investigate disc settling in the context of the downsizing picture, we compare the current stellar mass with respect to the bar age in Fig.~\ref{fig_barAge_mstar}. We use the stellar mass measurements from S$^4$G (\citealp{munoz2013impact}, \citeyear{munoz2015spitzer}), from $3.6~\mu\mathrm{m}$ fluxes. Interestingly, we do not find a correlation between the bar age and stellar mass. This indicates that some massive galaxies are still forming bars -- at least for the stellar mass regime of $\mathrm{M}_\star \geq 10^{10} \mathrm{M}_\odot$ --, in contrast with the downsizing predictions. However, this is in agreement with recent analysis presented in \cite{fragkoudi2024bar}, where the authors find that, for a sample from Auriga simulations (\citealp{grand2017auriga}, \citeyear{grand2019gas}) in a similar mass range, that despite the oldest bars are preferentially found in massive galaxies, there is not a correlation between the bar age and the stellar mass. Lastly, we also considered the stellar masses derived in \cite{querejeta2015spitzer} and \cite{leroy2019z} in Appendix~\ref{sec_appendixMass}, reaching similar results. 

We also investigate how different galaxy properties relate to bar ageing, and how these properties are affected by physical processes connected to bar-driven evolution. In Figure \ref{fig_barAge_nD_bar_A2} we show the normalized bar length and strength as a function of the bar age. We consider the bar lengths derived in \cite{herrera2015catalogue}, where the authors visually estimate the length of the bar from the distance in between the bar ends, and for bar strength, we consider the values from \cite{diaz-garcia2016}, as given by the $m$=2 Fourier mode of the galaxy surface brightness (A2). Both parameters are normalized to the size of the host galaxy, for which we consider the semi-major axis of the $25.5$ mag arcsec$^{-2}$ isophote, measured on the $3.6~\mu$m band by S$^4$G (\citealp{munoz2015spitzer}). Throughout this work, we bootstrapped our sample 1000 times to derive robust \textit{Pearson's} coefficients and $p-values$. We find a Pearson correlation coefficient of $r=0.654\pm0.004$, with $p-value=0.010\pm0.001$ for the normalized bar length with respect to bar age, and $r=0.480\pm0.005$ with $p-value=0.097\pm0.005$ for normalized strength with respect to bar age. 

\textcolor{black}{Following the same approach}, we investigate the nuclear disc size  (given in Table~\ref{table_sampleResults}) normalized by galaxy size with respect to bar age in Figure~\ref{fig_barAge_ND}, finding a correlation with Pearson coefficient index of $r=0.602\pm0.004$ and $p-value=0.025\pm0.003$. We also compare the nuclear disc size normalized by bar length with bar ages in Figure~\ref{fig_barAge_NDRbar}, where we find a constant size relation of $\sim0.12$, independent of bar age. This suggests that the nuclear disc is set by the size and growth of the bar. \textcolor{black}{Finally, we compare the relative nuclear disc stellar mass (i.e., normalized by the galaxy total stellar mass) with bar age in Figure~\ref{fig_barAge_NDmass}, finding a remarkable correlation, with $r=0.850\pm0.002$ and $p-value=2\times10^{-4}\pm0.7\times10^{-5}$. We derive the stellar masses of the nuclear discs by integrating their SFH (blue-dashed curve in Figs~\ref{fig_lSFmain} and \ref{fig_hSFmain}), following \cite{de2023new,de2023disc}, and present the values in Table~\ref{table_sampleResults}. Moreover, we colour-code Figure~\ref{fig_barAge_NDmass} by the median $\Sigma_\mathrm{SFR}$, where blue and red colours indicate higher and lower values of it, respectively. The mass build-up of the nuclear disc becomes clear, where young bars present less massive nuclear discs with higher star formation and, as the bar ages, the nuclear disc mass increases while star formation decreases. Our findings regarding the nuclear disc evolution with bar ageing are in good agreement with the inside-out scenario, proposed by \cite{bittner2020inside}, in which nuclear discs are built in rings of star formation that move outwards, growing and building up their mass. From these analyses, one notices that all the mentioned normalized properties are correlated with the bar age, where A2 shows the weakest trend -- note that when we normalize A2 by stellar mass, we find weaker or no correlation. Therefore, for the first time employing directly measured bar ages, we have observational indications that bars and nuclear discs can grow, build up their mass, and strengthen with time.}

To address whether or not bars are related to the host galaxy quenching, we compare the main-sequence offset ($\Delta\mathrm{MS}$ -- from \citealp{leroy2019z}) with bar ages in Figure~\ref{fig_barAge_quench}, available for $18$ galaxies in our sample. We find a considerable anti-correlation between $\Delta\mathrm{MS}$ and bar ageing, with $r=-0.699\pm0.0003$ and $p-value=0.007\pm0.001$, suggesting the importance of the bar in suppressing the star formation of the host galaxy. In Figures \ref{fig_barAge_dSFR} and \ref{fig_barAge_mHI}, we analyse the median $\Sigma_\mathrm{SFR}$ within the nuclear disc and the neutral hydrogen supply of the host galaxy ($\mathrm{M}_\mathrm{HI}$) with respect to bar ages, investigating secondary consequences of galaxy quenching. For the former, we also find a strong anti-correlation with a Pearson correlation coefficient of $r=-0.791\pm0.003$, with $p-value=0.0012\pm0.0006$; while for last, we find a weak trend, with values of $r=-0.463\pm0.007$ and $p-value\sim0.0151\pm0.007$. These results agree with the scenario in which the bar funnels gas inwards and, with the decrease of available gas, the nuclear disc also tends to form fewer stars. However, the bar does not completely exhaust the gas present in the galaxy. 

%##################################################################%
%##################################################################%

\section{Discussion}
\label{sec_Discussion}

\subsection{When do galactic discs settle and bars form?}

We derive the bar formation epochs for 18 galaxies in the TIMER survey, forming the largest sample of nearby galaxies with known bar ages. Including the galaxies from \cite{de2023disc}, we find ages between $\sim1-13.0~\mathrm{Gyr}$, which corresponds to redshifts between $\sim0-6$ (nonetheless, note older stellar populations have higher intrinsic uncertainties). Since numerical and theoretical work suggest that galaxies can form a bar once their discs are dynamically mature (at least to a significant extent; \citealp[e.g.,][]{kraljic2012two} and references therein), our result implies that the necessary conditions to form bars are already in place for some galaxies at those early phases of the evolution of the Universe. This would generally be interpreted as indicating that self-gravitating disc galaxies, baryon-dominated, with relatively low velocity-dispersion -- where rotational motions are more significant than pressure-supported motions -- exist since $z\geq2$. This is in agreement with recent observational findings from ALMA (e.g, \citealp{smit2018rotation}; \citealp{neeleman2020cold}; \citealp{rizzo2020dynamically}; \citealp{lelli2021massive}; \citealp{posses2023structure}; \citealp{lelli2023cold}) and JWST (e.g., \citealp{ferreira2022panic}, \citeyear{ferreira2023jwst}; \citealp{nelson2022jwst}; \citealp{jacobs2023early}). Furthermore, we find galaxies that formed bars relatively recently, with ages below $5~\mathrm{Gyr}$ ($z \leq 0.5$), indicating that bar formation is an ongoing process in the Universe. In other words, even though currently bars are understood as old structures -- which are robust and long-lived (e.g., \citealp{athanassoula2003angular}, \citeyear{athanassoula2005nature};  \citealp{gadotti2015muse}; \citealp{perez2017observational}; \citealp{de2019clocking}; \citealp{rosas2020buildup}; \citealp{fragkoudi2020chemodynamics}; \citealp{de2023new,fragkoudi2024bar}; \citealp{verwilghen2024simulating}), some bars can still be young and recently formed (see also \citealp{de2023disc}; \citealp{fragkoudi2024bar}). Whether or not the young bars in our sample are first-generation or reformed bars is a discussion beyond the scope of this work.

Even though some studies find that high redshift discs are often thick and turbulent (e.g., \citealp{elmegreen2006observations}; \citealp{newman2013sins}), which indicates that these objects are still under great influence of external processes such as minor mergers, recent works find that this can depend on the choice of gas tracer (e.g., \citealp{rizzo2024alma}). Other studies which investigate galaxy morphology at high redshifts ($z\geq1.5-4.0$) find the presence of well-developed disc galaxies (e.g., \citealp{shapiro2008kinemetry}; \citealp{schreiber2009sins}; \citealp{epinat2012massiv}; \citealp{wisnioski2015kmos3d}; \citealp{rizzo2020dynamically}; \citealp{lelli2021massive}; \citealp{posses2023structure}). More recently, works using JWST data find that disc galaxies, with significant rotational support, can be the majority up to $z\sim8$ (e.g., \citealp{ferreira2022panic}, \citeyear{ferreira2023jwst}; \citealp{nelson2022jwst}; \citealp{jacobs2023early}), in agreement with our results that disc galaxies with relatively cold dynamics exist since the Universe was $\sim 1-2~\mathrm{Gyr}$ old. Even though we present a sample of only $20$ galaxies, this work brings a novel and independent benchmark of when internal, secular evolution started to take place in our sample. %Nevertheless, we would like to point out that our sample is morphologically biased and limited to relatively massive galaxies, and a larger volume-limited sample is necessary to constrain further when this transition happens in the Universe. 

Some simulations that investigate how the bar fraction evolves with time find that bars can exist since $z > 2$ (e.g., \citealp{kraljic2012two}; \citealp{rosasguevaraetal2020}; \citealp{fragkoudi2020chemodynamics}, \citeyear{fragkoudi2024bar}), which is also in good agreement with our results. More specifically, \cite{rosasguevaraetal2020} find, for an IllustrisTNG sample, that $30\%$ of the galaxies are barred at $z\approx6$ and, while their results have a discrepancy with observational data at intermediate redshifts, they argue that this discrepancy could be a result of observational detection limitations. On the other hand, cosmological zoom-in simulations (e.g. \citealt{kraljic2012two,fragkoudi2024bar}) tend to find a decreasing bar fraction with redshift, in agreement with our results. On the observational side, \cite{le2024jwst} derived bar fractions for a sample at $1\leq z \leq 3$, based on JWST and HST data separately, finding that the bar fraction is about twice as large with JWST data, as compared to HST data. This illustrates the limitations concerning the detection of bars in previous works at higher redshifts. For the JWST data -- better suited for higher redshifts -- \citeauthor{le2024jwst} find bar fractions of  $14\%$ in bins between $1\leq z \leq 3$. In parallel, recent work by \cite{guo2024abundance} find a bar fraction of $2.4-6.4\%$ at  $3\leq z \leq 4$.

\begin{figure*}
\centering
\includegraphics[width=\linewidth]{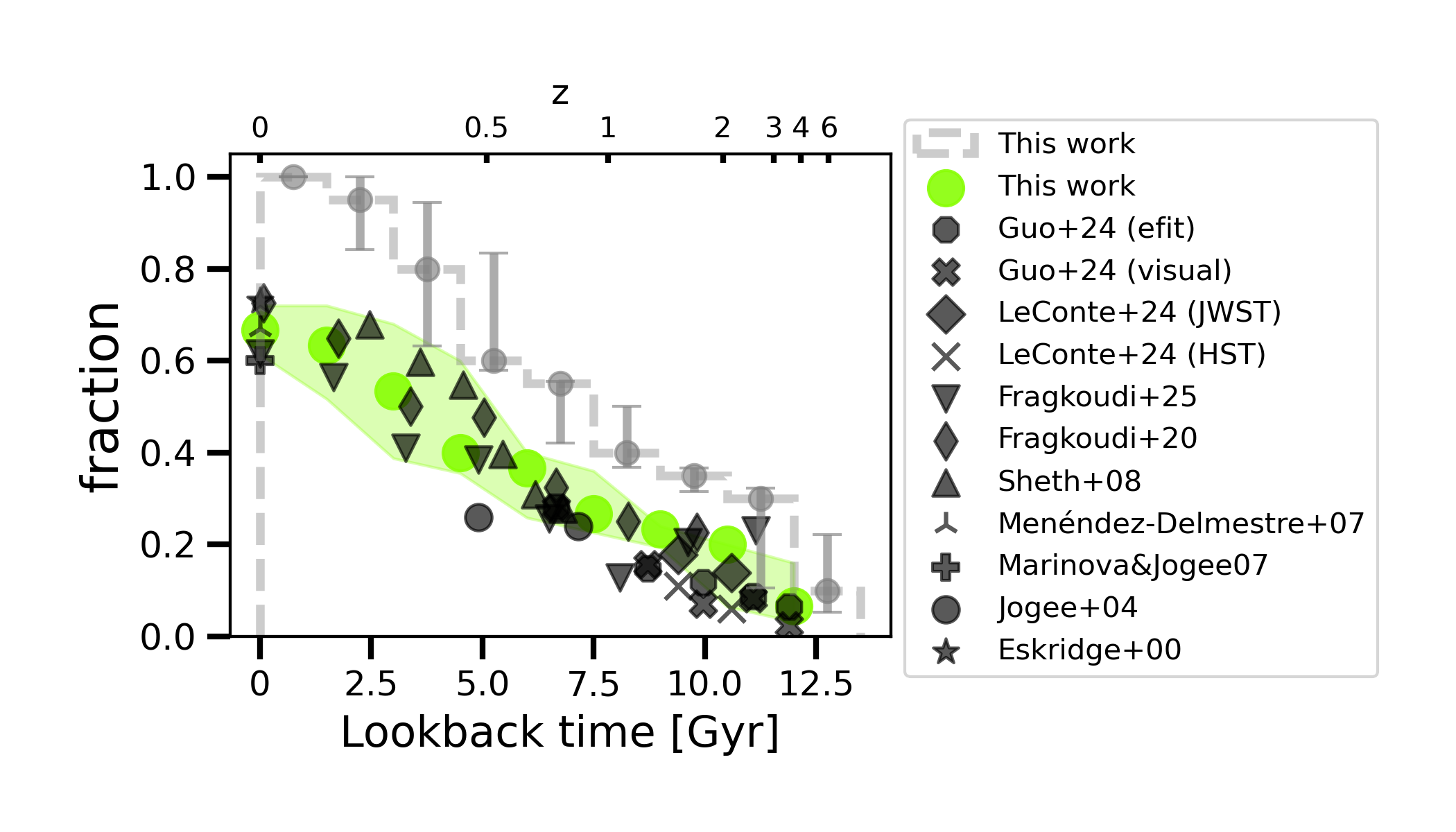}
    \caption{\textbf{Extrapolated bar fractions for different redshifts from this work.} Based on Figure~\ref{fig_histBarAges}, we display our cumulative distribution (dashed-gray histogram) and normalize it by the observed bar fraction in the Local Universe in studies that consider both weak and strong bars ($0.67$ -- e.g., \citealp{eskridge2000frequency}; \citealp{menendez2007near}; \citealp{marinova2007characterizing}). With this, we derive the extrapolated bar fraction over time in our sample (green dots). Additionally, the shaded green region accounts for the systematic error in our bar age measurements of $^{+1.8}_{-1.4}~\mathrm{Gyr}$. With the grey error bars, we also display the likely ranges in the cumulative distribution considering this systematic error as well. We compare our results with the observed bar fraction from different works, finding a remarkable agreement with works that consider both weak and strong bars.}
    \label{fig_histBarAgesFractions}
\end{figure*}

To understand how our bar age estimates compare to the observed evolution of the bar fraction over time, in Fig.~\ref{fig_histBarAgesFractions} we derive the bar fraction in our sample across cosmic time directly from our derived bar ages, as follows. We compute the cumulative fraction of barred galaxies as a function of lookback time from our bar ages, assuming that once a bar is formed, it is not destroyed. However, since all of our galaxies are barred, we would thus find a bar fraction of 100$\%$ in the Local Universe. To rectify this offset from the observed bar fraction a $z=0$, we normalise it to the observed bar fraction of $67\%$ (e.g., \citealp{eskridge2000frequency}; \citealp{menendez2007near}; \citealp{marinova2007characterizing}), and compute the rest of the cumulative distribution accordingly. These extrapolated bar fractions we measure can be considered as a lower limit only due to two main caveats: (\textit{i}) we only consider galaxies that are barred today, and (\textit{ii}) we assume that all galaxies in our sample had a clear disc morphology at all redshifts considered. In the case of (\textit{i}), it is possible that there were more bars in the past that were eventually destroyed (e.g., \citealp{rosasguevaraetal2020}), which would increase the bar fraction at early epochs. Considering the assumption in (\textit{ii}), it is possible that some galaxies in our sample hosting younger bars would only develop a clear disc galaxy structure later in time, which would decrease the number of disc galaxies and thus increase the fraction of barred galaxies at earlier times. Additionally, it is expected that disc galaxies can suffer a morphological transformation, and no longer be considered in these fraction measurements, which can not be accounted for here. Despite these caveats, we find in Fig.~\ref{fig_histBarAgesFractions} a remarkable agreement between our extrapolated bar fractions and the bar fractions at different redshifts -- from both observations and zoom-in cosmological simulations --, especially with works that consider both strong and weak bars. We want to emphasize that this is a novel method to investigate bar fractions over time, which is also a completely independent approach to the matter, with a starting point in the derived bar ages in nearby galaxies.

\subsection{Bar formation in the downsizing scenario}

When it comes to disc settling and bar formation, the downsizing scenario (e.g., \citealp{cowie1996new}; \citealp{thomas2010environment}; \citealp{sheth2012hot}) predicts that more massive galaxies would host older bars. In other words, massive galaxies in the Local Universe would have achieved enough mass to dynamically settle their discs first, hence forming their bars first. To test this scenario, we assessed the relation between bar age and the current stellar mass in Fig.~\ref{fig_barAge_mstar}, finding no correlation in the mass range of our sample, \st{in contradiction} {which does not agree with the downsizing scenario predictions. More specifically, we find young bars ($\sim4~\mathrm{Gyr}$) with stellar masses ranging between $2 \leq \mathrm{M}_\star \leq 18 \times 10^{10} \mathrm{M}_\odot$, which indicates that even some massive galaxies formed their bars more recently, in accordance with \cite{de2023disc}. However, we can not yet diagnose if the young bars in massive galaxies are first-generation bars or reformed bars. More recent simulations find that bars formed after $z\leq2$ tend to be robust and long-lived (e.g., \citealp{rosas2020buildup}; \citealp{fragkoudi2024bar}; \citealp{verwilghen2024simulating}; \citealp{rosas2024rise}). \cite{rosas2024rise} demonstrate for a TNG50 sample that bars are mostly stable and, if they are destroyed after $z\sim1$, this is usually due to mergers and the environment. However, only $\sim6\%$ of the barred galaxies at $z=1$ have a barless disc-like morphology at $z=0$. In summary, the destruction of bars usually takes place before $z\sim1-2$, and we do not discard this possibility in the case of our young bars.

\cite{erwin2018dependence} investigates the bar fraction for different masses in a S$^4$G sample of late-type galaxies (\citealp{sheth2010spitzer}), finding that the fraction of barred galaxies decreases for galaxies with masses greater than $10^{10}\mathrm{M}_\odot$, which is the TIMER sample regime. Considering bars are robust structures in the context of the downsizing prediction, one would expect that massive galaxies would form bars first and sustain them until $z=0$. In other words, the fraction of barred galaxies would increase with mass, which is not observed in \cite{erwin2018dependence} -- although the author does not include lenticular galaxies in his sample. This is consistent with our main results, in which we find that the downsizing scenario might not be sufficient to {\em completely} explain bar formation, \color{black} at least in the mass regime we probe in this work (a sample including lower mass galaxies would be beneficial in this context). \color{black} 

Even though the build-up of stellar mass contributes to the `dynamical maturity' of galaxies, other processes will also play a role in the formation of bars. For example, theoretical studies have shown that when a galaxy is less disc-dominated (i.e. when the dark matter dominates over the stellar disc in the central regions), or if the disc is kinematically hotter, this will contribute to delaying or even suppressing bar formation (e.g. \citealt{hohl1971numerical, lynden1972generating, ostriker1973numerical,athanassoula1986bi,combes1990box,berentzen2007gas,bland2023rapid, ghosh2023bars,fragkoudi2024bar}). Furthermore, studies find that having a gas fraction in the disc above $10\%$ might suppress or delay bar formation (e.g. \citealt{athanassoula2013bar}). Recently, \cite{fragkoudi2024bar} showed that, in cosmological simulations, the time of bar formation is well correlated with disc dominance at $z=0$, as well as with the time at which the galaxy becomes disc-dominated. While there is a correlation between stellar mass and how disc-dominated (or baryon-dominated) a galaxy is\footnote{The stellar fraction peaks around the mass of the Milky Way according to abundance matching relations (e.g. \citealt{moster2013}).}, the fact that we do not find a clear correlation between mass and bar age neither in cosmological simulations (e.g. \citealt{fragkoudi2024bar}) nor in this work, suggests that stellar mass is not enough to constrain the time of bar formation. Furthermore, interactions may destroy or induce the formation of a bar -- even in an otherwise stable disc --  depending on the characteristics of the involved galaxies and the orbital properties of the interaction (e.g., \citealp{noguchi1987close}; \citealp{gerin1990influence}; \citealp{gadotti2009structural}; \citealp{li2009clustering}; \citealp{mendez2012nature}; \citealp{lokas2014adventures}; \citealp{mendez2023jwst}). This could be the reason for the relatively young bar we find in NGC\,1097, the most massive galaxy in our sample. This adds complexities to the downsizing picture, which is not necessarily ruled out by our results.

In addition to the bar formation itself, the expected disc settling moment for different masses is also an open question. Many different works investigate which physical processes are responsible for the disc formation, that is, the transition between a disordered motion supported galaxy to a rotationally supported low-dispersion object (e.g., \citealp{fall1980formation}; \citealp{ryden1987galaxy}; \citealp{okamoto2005effects}; \citealp{brook2012thin}; \citealp{christensen2016n}; \citealp{stern2021virialization}; \citealp{conroy2022birth}; \citealp{hafen2022hot}; \citealp{gurvich2023rapid}; \citealp{semenov2024formationI}, \citeyear{semenov2024formation}). \cite{semenov2024formationI} demonstrates that the disc formation time can depend on how the halos assembled mass, and if destructive mergers took place late on, which would reset the disc formation process. For example, \cite{fragkoudi2024bar} showed that Milky Way mass disc galaxies that host bars tend to assemble their stellar mass earlier on than unbarred disc galaxies. \cite{rosas2020buildup} argue that for an IllustrisTNG Milky-Way-like sample, late disc formation is associated with mass assembly histories with significant mergers taking place later. In summary, the scenario of disc formation and settling is complex and remains an open question, but it does not seem to necessarily relate only to the mass of the galaxy. %To investigate further the downsizing relation with disc settling, we need to extend our analysis to a volume-limited sample that includes galaxies with lower masses.

\subsection{How do bars grow?}
\label{subsection_barGrowth}
For the first time, it is possible to analyze how different properties of galaxies change with the ageing of bars, and how these observational results compare to theoretical work. 

When it comes to bar evolution, especially bar growth, many studies reach different conclusions. From an observational viewpoint, \cite{kim2021cosmic} analyzed the data from the Cosmic Evolution Survey (COSMOS -- \citealp{scoville2007cosmos}; \citealp{koekemoer2007cosmos}) and found that the average bar length does not show signs of evolution in the past $7~\mathrm{Gyr}$ -- both in normalised and absolute terms. This does not agree with the findings from \cite{rosas2022evolution}, where the authors found, for an IllustrisTNG sample (\citealp{pillepich2018first}; \citealp{nelson2018first}), that bars do grow in absolute terms. However, the growth takes place at a similar pace as the disc growth, maintaining a fairly constant size relation. In agreement, \cite{zhao2020barred} also found for an IllustrisTNG sample that, in the past $6~\mathrm{Gyr}$, the absolute length of bars increases by a factor of $0.17~\mathrm{dex}$. Furthermore, \cite{anderson2022secular} and \cite{erwin2023profiles} argue that the presence of ``shoulders'' -- one type of bar profile structure -- is evidence of secular bar growth. This is a brief illustration of the lack of agreement between observations and simulations on whether or not bars grow in length as they evolve, both in absolute terms or normalised.

Employing our derived bar ages, we can investigate the normalized bar size evolution as in Fig.~\ref{fig_barAge_nD_bar_A2} and find an indication that younger bars are smaller than older bars when compared to the size of the galaxy. This is consistent with the picture in which the normalized length of bars can increase with their ageing. However, \cite{fragkoudi2024bar} showed that the picture is not that simple. For an Auriga simulation sample (\citealp{grand2017auriga}, \citeyear{grand2019gas}) of galaxies, they retrieved the same relation between normalized bar length with bar age that we find in this work. Still, investigating further the evolution of said bars, they found that not every bar has grown with time. The authors demonstrated that bars which formed at $z\geq 2$ already form large and do not grow, while bars that formed at intermediate redshifts ($z\leq 2$) do grow with time. In other words, the authors retrieve from their simulations the same relation presented here, but in a scenario in which not all bars have grown.

\subsection{Bar-driven galaxy quenching}

Several theoretical, numerical, and observational works find that bars, once formed, affect their host galaxy and drive its evolution (e.g.,
\citealp{masters2012galaxy}; \citealp{schawinski2014green}; \citealp{haywood2016milky}; 
\citealp{fragkoudi2020chemodynamics}; \citealp{geron2021galaxy}; \citealp{rosas2022evolution}). When comparing a barred sample with a counterpart unbarred sample from the SDSS-IV MANGA survey, \cite{fraser2020sdss} found that barred galaxies peak their star formation history, quench, and form most of their stellar mass earlier than the unbarred ones. One important process of bar-driven secular evolution is the funnelling of gas towards the central region, forming the nuclear disc and often causing suppression of star formation along most of the bar (e.g., \citealp{masters2012galaxy}; \citealp{schawinski2014green}; \citealp{geron2021galaxy}).

We consider the $\Delta\mathrm{MS}$ measurements for $18$ of the galaxies in our sample (see \textcolor{black}{Figs.~\ref{fig_barAge_NDmass} and} \ref{fig_barAge_quench}), finding a strong anti-correlation with bar age. That is, galaxies with older bars are more `quenched' than galaxies with younger bars, with the SFR in the nuclear disc following the same relation (Fig.~\ref{fig_barAge_dSFR}). Interestingly, when we analyse the availability of neutral gas with bar ages, the trend is weaker and not statistically significant (see Fig.~\ref{fig_barAge_mHI}). This indicates that, even though the bar can aid in the suppression of star formation in the main disc, it does not exhaust the gas, suggesting the gas is less efficient in forming stars (e.g., \citealp{saintonge2016molecular}; \citealp{bacchini2019volumetric}; \citealp{pessa2022variations}). Further investigation of the star formation efficiency is beyond the scope of this work.

We find interesting correlations between quenching and bar ages in this work. Nevertheless, as illustrated in \cite{fraser2020sdss}, even though the bar can aid in the suppression of star formation in the galaxy, it is not clear what took place first: the quenching of the galaxy or the formation of the bar -- since bars tend to form in galaxies with less gas (e.g., \citealp{berentzen2007gas}; \citealp{athanassoula2013bar}). To properly confirm this picture, in the near future, we plan to investigate the gas and star formation rate properties in the entire galaxy in detail and break this degeneracy for the first time.

\subsection{Double-barred galaxies are older systems}

Studies report that $12-30\%$ of barred galaxies host a second, inner/nuclear bar (e.g., \citealp{erwin2004double}; \citealp{buta2015classical}), with sizes varying between $0.3-2.5~\mathrm{kpc}$ (\citealp{de2020deconstructing}). Historically, mainly two formation scenarios for the inner/nuclear bar have been considered: (\textit{i}) they are formed from the gas brought inwards from the main bar and/or mergers and interactions, firstly forming a gaseous bar, which forms stars and give place to a fragile stellar bar (e.g., \citealp{friedli1993secular}; \citealp{heller1994fueling}); or, (\textit{ii}) they are formed dynamically from instabilities in the nuclear disc, similar to the main bar (e.g., \citealp{debattista2006long}; \citealp{du2015forming}). In the first scenario, inner/nuclear bars are young and transient structures, being destroyed after just a few hundred Myr. On the other hand, in the second scenario, the inner/nuclear bars share similarities with the main bar (e.g., \citealp{de2019clocking}; \citealp{bittner2021galaxies}), being long-lived structures with evidence of buckling in some cases (e.g., \citealp{mendez2019inner}). In fact, \cite{bittner2021galaxies} argues that double-barred systems behave as ``galaxies inside galaxies''. Beyond these two scenarios, \cite{semczuk2024new} proposed that tidal interactions can trigger the formation of the inner/nuclear bar before the outer bar in some cases.

In \cite{de2019clocking}, the authors demonstrated that, between the two main scenarios, the second scenario is more likely for NGC\,1291 and NGC\,5850, measuring the formation epoch of the main bars to be above $6.5$ and $4.5~\mathrm{Gyr}$, respectively -- in accordance with the results presented here. Furthermore, other efforts from the TIMER collaboration (e.g., \citealp{mendez2019inner}; \citealp{bittner2021galaxies}) find that inner/nuclear bars are formed through dynamical instabilities in the nuclear discs and that inner/nuclear bars are long-lived systems.} \color{black} These findings suggest that the second scenario for the formation of the inner/nuclear bar is more likely. \color{black} Nevertheless, using N-body/hydrodynamical simulations, \cite{wozniak2015can} finds that it is also necessary to have a gaseous component with star formation in order to maintain the inner/nuclear bar long-lived and stable.\color{black}

As discussed in Section~\ref{subsubsec_MultipleNuclearComponents}, four of the galaxies in our sample have been reported to host an inner/nuclear bar, which are among the oldest bars of our sample. Even though we do not estimate the formation epoch of the inner/nuclear bars, our results agree with the scenario in which inner/nuclear bars form from disc instabilities in the nuclear disc, which require building the mass of the nuclear disc itself, hence longer timescales and older systems -- with bar ages $>7.5~\mathrm{Gyr}$. Additionally, studies show that a high gas fraction on the main disc can delay the formation of the main bar (e.g., \citealp{berentzen2007gas}; \citealp{athanassoula2013bar}; \citealp{seo2019effects}). If this is the case also for the inner/nuclear bars, it would be necessary to quench the star formation in the nuclear disc, or at least to consume enough of the gas, to make it possible for the inner/nuclear bar to form. Consistent with that, three of the four doubled-barred systems in our sample belong to the low-SF sub-sample. Lastly, since the investigation of these systems using spatially resolved spectroscopy is very recent, there are few constraints on the timescales for their formation (e.g., \citealp{de2019clocking}). Overcoming the physical limitations of identifying and studying inner/nuclear bars is a challenge, and further efforts with larger samples are necessary.
%##################################################################%
%##################################################################%

\section{Summary and conclusions}
\label{sec_Summary}

Studies suggest that, in a simplified picture, disc galaxies evolve in a two-phase scenario: at first, the main drivers of galaxy evolution are external, such as mergers and interactions; later, with the expansion of the Universe and the galaxy assembling enough mass, internal evolutionary processes start to play an important role - often referred to as ``secular evolution''. The epoch in which this transition happens is a long-standing question. One approach to understanding this transition in individual objects is to time when stellar bars appeared since they can only form once the disc is dynamically mature, at least partially (e.g., \citealp{kraljic2012two} and references within). 

We summarize this work as follows:

\begin{itemize}
    \item We have derived, for the first time, the ages of bars in a sample of $20$ galaxies mostly from the TIMER survey (\citealp{gadotti2019time}), using the methodology presented in \cite{de2023new};
    \item We find bar ages between $1-13~\mathrm{Gyr}$ (Table~\ref{table_sampleResults}, Figs.~\ref{fig_histBarAges}, \ref{fig_lSFmain}, and \ref{fig_hSFmain}), illustrating how settled discs exist in the Universe at least since $z>2$, and that the dynamical maturing of discs is an ongoing process;
    \item We find strong anti-correlations between bar ages and star formation rate surface density in the nuclear disc and main sequence offset $\Delta\mathrm{MS}$ for the entire galaxy (Fig.~\ref{fig_barAge_quench}), in accordance with the scenario in which the presence of the bar aids the suppression of star formation in the galaxy. However, the neutral gas supply is only weakly anti-correlated with bar ages, suggesting less efficient star formation rather than gas exhaustion;
    \item Separating our sample according to the star formation rate surface density in the nuclear disc region, into low- and high-SF nuclear discs, we find that the former are hosted by older bars while the latter are hosted by younger bars, with mean ages of $9.3\pm3.6~\mathrm{Gyr}$ and $4.0\pm2.0~\mathrm{Gyr}$, respectively. This is in agreement with the scenario in which bars swipe the gas of the disc towards the central region, aiding the suppression of star formation in the main disc;
    \item Contrary to predictions from the downsizing picture, we find no correlation between bar age and current galaxy stellar mass (Fig.~\ref{fig_barAge_mstar}), indicating that, even though having enough mass is necessary to form bars, it is not enough to explain all bar formation. In other words, other factors, such as galaxy interactions and/or gas fractions and disc-dominance, play an important role in determining if and when a bar forms in a given disc galaxy, \color{black} at least in the mass regime we probe\color{black};
    \item We find correlations between bar ages and normalized bar length and strength, suggesting the secular evolution of these structures, whereby older bars tend to be larger and stronger (but see also discussion in Sec.~\ref{subsection_barGrowth});
    \item We find a strong correlation between bar ages and nuclear disc size normalized by galaxy size, whereby older bars host relatively larger nuclear discs. In line with this, we find that the nuclear disc size normalized by bar {\em length} does not depend on the bar ageing. \textcolor{black}{Additionally, the nuclear disc stellar mass is tightly correlated with the bar age, where older bars host more massive nuclear discs with less star formation, and young bars host less massive nuclear discs with higher star formation}. These findings suggest the co-evolution of the bar and nuclear disc, with the nuclear disc following the inside-out scenario proposed in \cite{bittner2020inside}, with a dependency on bar properties, such as length; 
    \item From our bar ages we estimate how the bar fraction evolves over time (Fig.~\ref{fig_histBarAgesFractions}), finding remarkable agreement with observed bar fractions in different redshifts with a completely independent and novel approach;
    \item We find that systems with double bars tend to be among the oldest in our sample, and have low star formation rate surface density. This is in agreement with the scenario in which the inner/nuclear bar is formed from disc instabilities in the nuclear disc, which generally takes relatively longer time scales.
\end{itemize}

In this work, we provide, for the first time, an observational estimate of the bar ages for a sample of nearby galaxies. This provides us with an independent insight into the cosmic epoch in which disc galaxies started to dynamically mature, and when bar-driven, secular evolution processes started to take place in the Universe and shape the properties of disc galaxies. Our results are in agreement with high-redshift observational studies on the bar fraction beyond redshift one and up to redshift four. In addition, we are able the obtain insights on secular evolution directly connected to the measured bar ages, finding evidence on the growth of bars and bar-built nuclear discs, and on the role of bars in quenching star formation in their host galaxies. 

%##################################################################%
%##################################################################%

\begin{acknowledgements}
We acknowledge the insightful comments of the referee, Dr. H. Wozniak, which improved and clarified the work presented here. Based on observations collected at the European Organisation for Astronomical Research in the Southern Hemisphere
under ESO programmes 097.B-0640(A), 095.B-0532(A), and 060.A-9313(A), 096.B-0309, and 0100.B-0116. All raw and reduced data are available in the ESO Science Archive Facility. This work was supported by STFC grants ST/T000244/1 and ST/X001075/1. T.K. acknowledges support from the Basic Science Research Program through the National Research Foundation of Korea (NRF) funded by the Ministry of Education (No. 2022R1A4A3031306, No. RS-2023-00240212). J.F-B acknowledges support from the PID2022-140869NB-I00 grant from the Spanish Ministry of Science and Innovation. AdLC acknowledges financial support from the Spanish Ministry of Science and Innovation (MICINN) through RYC2022-035838-I and PID2021-128131NB-I00 (CoBEARD project). PSB acknowledges support from grant PID2022-138855NB-C31, funded by the Spanish Ministry of Science and Innovation, MCIN/AEI/10.13039/501100011033/FEDER, EU. MQ acknowledges support from the Spanish grant PID2022-138560NB-I00, funded by MCIN/AEI/10.13039/501100011033/FEDER, EU. JMA acknowledges the support of the Viera y Clavijo Senior program funded by ACIISI and ULL and the support of the Agencia Estatal de Investigación del Ministerio de Ciencia e Innovación (MCIN/AEI/10.13039/501100011033) under grant nos. PID2021-128131NB-I00 and CNS2022-135482 and the European Regional Development Fund (ERDF) ‘A way of making Europe’ and the ‘NextGenerationEU/PRTR’. PC acknowledges support from Conselho Nacional de Desenvolvimento Científico e Tecnológico (CNPq) under grant 310555/2021-3 and from Fundação de Amparo à Pesquisa do Estado de São Paulo (FAPESP) process number 2021/08813-7. 
\end{acknowledgements}

%##################################################################%
%##################################################################%

\bibliographystyle{aa} % style aa.bst
\bibliography{bibliography} % your references Yourfile.bib

\begin{appendix}
\section{Different mass measurements}
\label{sec_appendixMass}

\begin{figure}[hbt!]
\centering
\includegraphics[width=\linewidth]{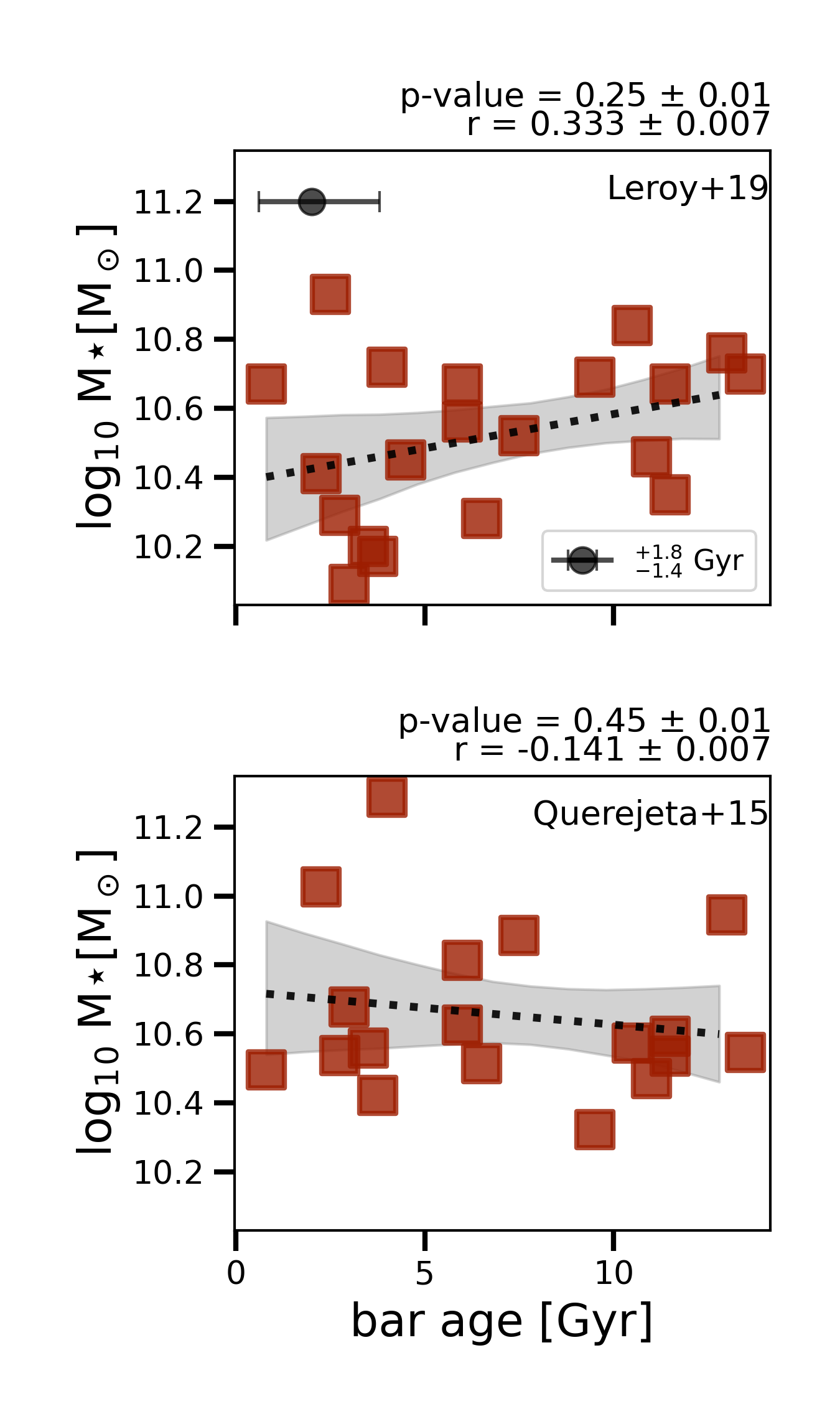}
    \caption{\textbf{Downsizing and bar formation for different mass measurements.} As in Fig. \ref{fig_barAge_mstar}, we compare the stellar mass of the host galaxy with respect to bar ages, considering values from \cite{leroy2019z} on the upper panel, and \cite{querejeta2015spitzer} on the bottom panel. For all three different mass measurements we consider in this work, we find no significant correlation between stellar mass and bar ageing.}
    \label{fig_barAge_mstarAppendix}
\end{figure}

In order to investigate the role of downsizing as a constraint to bar formation, we compare stellar mass with bar ageing, finding no correlation in Fig. \ref{fig_barAge_mstar}. To ensure the robustness of our results, we also consider mass measurements from two different works in Figure \ref{fig_barAge_mstarAppendix}: \cite{querejeta2015spitzer} and \cite{leroy2019z}. In \cite{querejeta2015spitzer}, the authors also consider Spitzer fluxes in the $3.6\mathrm{m}\mu$ band but apply a dust correction. On the other hand, \cite{leroy2019z} derive masses from mid-IR fluxes, either from IRAC1 or WISE1. As in Fig. \ref{fig_barAge_mstar}, we find no significant correlation between stellar mass and bar age, indicating that the downsizing picture of bar formation cannot completely explain the formation or the absence of bars in disc galaxies.

\section{Individual results}
\label{sec_appendix}

\begin{figure*}
\centering
\includegraphics[trim={0 6cm 0 0},clip,width=\linewidth]{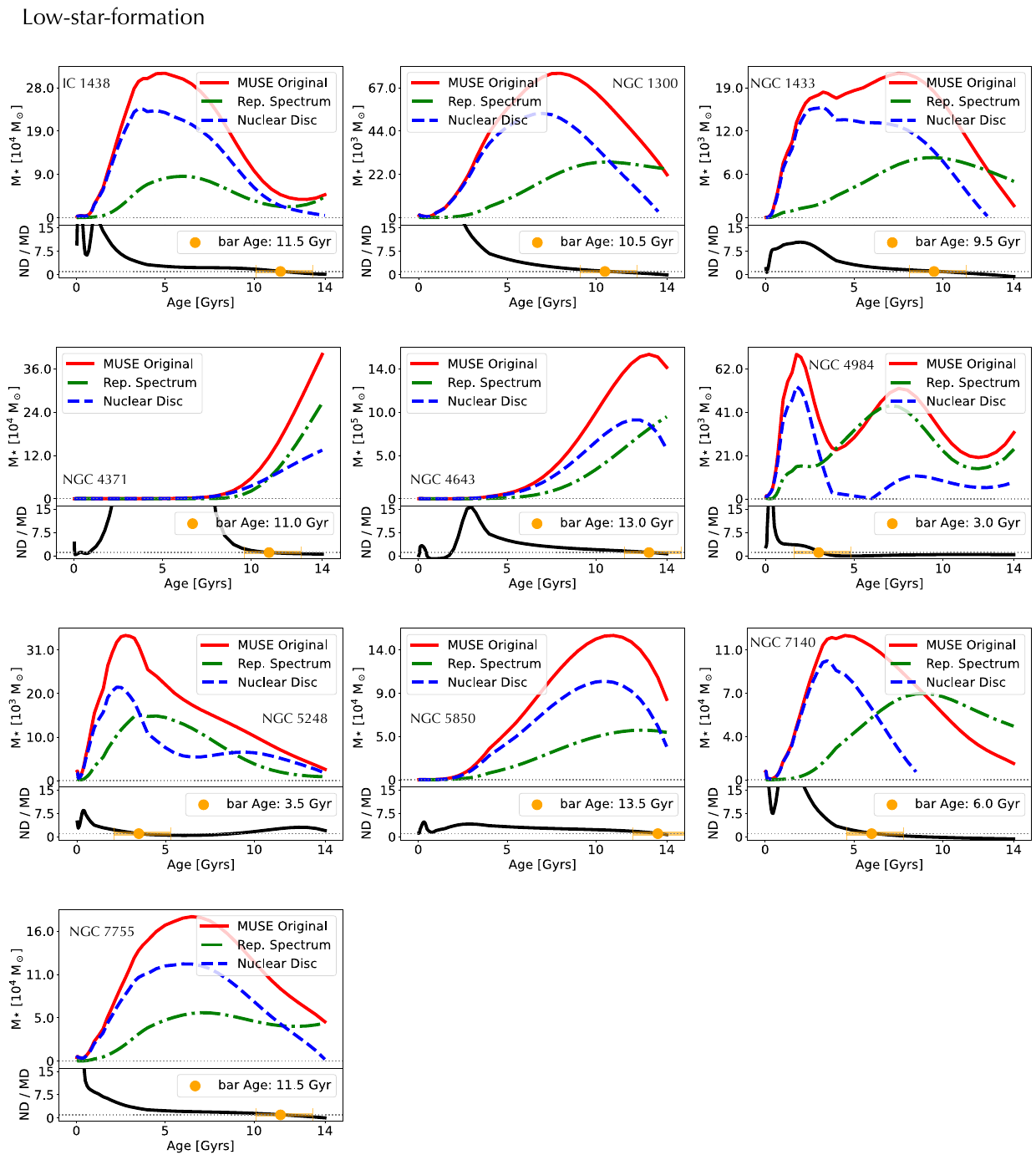}
    \caption{Individual measurements of bar age of the low-SF sample. Following the \cite{de2023new} methodology, we define the bar age as the moment in which the SFH of the nuclear disc (dashed-blue line) overcome the SFH of the main disc (dot-dashed-green line). Additionally, based on the tests performed in \cite{de2023new} and here, we estimate the systematic measurement error of $^{+1.8}_{-1.4}~\mathrm{Gyr}$. }
    \label{fig_lSFmain}
\end{figure*}

\begin{figure*}
\centering
\includegraphics[trim={0 12cm 0 0},clip,width=\linewidth]{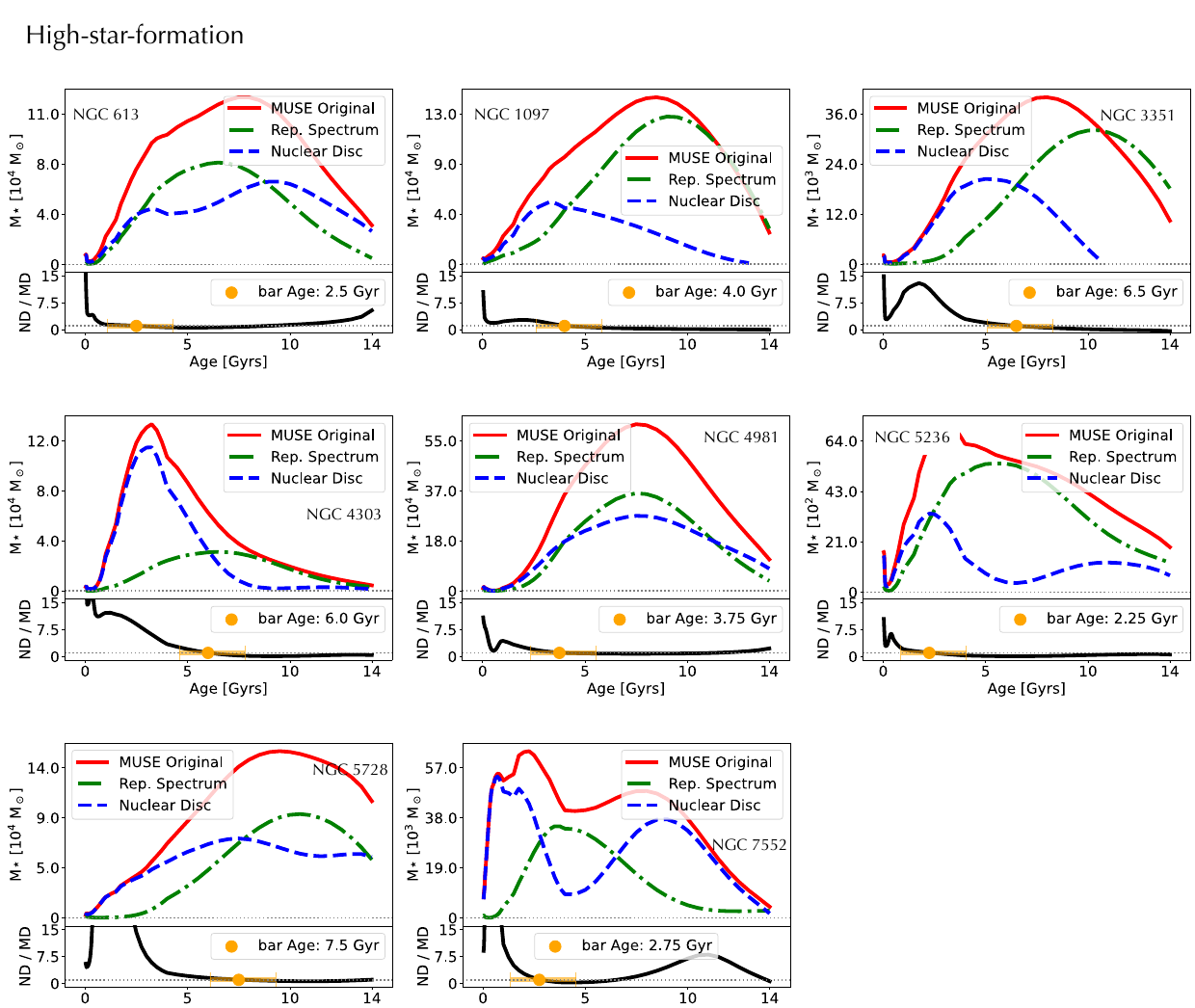}
    \caption{Same as Fig.~\ref{fig_lSFmain}, for the high-SF sub-sample.}
    \label{fig_hSFmain}
\end{figure*}

In this work, we derived individual bar ages for $18$ nearby galaxies from the TIMER sample (\citealp{gadotti2019time}), applying the methodology described in Section~\ref{sec_Methodology}. In this Appendix, we display the individual results in Figures~\ref{fig_lSFmain} and \ref{fig_hSFmain}, separated by the sub-samples of low- and high-SF, respectively. For each galaxy, we display the SFHs of the original datacube (solid-red curves), the modelled main disc (dot-dashed-green curves), and the clean nuclear disc (dashed-blue curves). Additionally, we display the ratio between the SFHs of the main disc and the nuclear disc at the bottom of each individual result. We highlight the region in which the criteria for bar age measurement are achieved in orange -- that is, ND/MD > 1 for the first time towards younger ages. Lastly, we also display the considered systematic errors of $^{+1.8}_{-1.4}~\mathrm{Gyr}$.

\end{appendix}
%##################################################################%
%##################################################################%
\end{document}